\documentclass{article}
\usepackage{amsfonts}
\usepackage{epsfig}
\begin{document}

\title{Homogeneous and Scalable Gene Expression \\ 
Regulatory Networks  with Random Layouts \\ 
 of Switching  Parameters }

\vspace{1cm}

\author{ {D. Volchenkov} \footnote{Visiter in the Centre de Physique Theorique; E-mail: dima427@yahoo.com}{\ } and R. Lima
\vspace{0.5cm}\\
{\it  Centre de Physique Theorique, CNRS, Luminy Case 907,}\\
{\it 13288,  Marseille CEDEX 09  France}}

\date{\today}
\maketitle
\large  

\begin{abstract}
\noindent

We consider a model of large regulatory gene expression networks
where the thresholds activating the sigmoidal interactions between genes and the signs of these interactions are shuffled randomly. Such an approach allows for a qualitative 
understanding of network dynamics in a lack of empirical data concerning the large genomes of living organisms.

Local dynamics of network nodes exhibits the multistationarity and oscillations and depends crucially upon the global topology of a "maximal" graph (comprising of all possible interactions between genes in the network). The long time behavior observed in the network defined on the homogeneous "maximal" graphs is featured by the fraction of positive interactions ($0\leq \eta\leq 1$) allowed  between genes. There exists a critical value $\eta_c<1$ such that if $\eta<\eta_c$, the oscillations persist in the system, otherwise, when $\eta>\eta_c,$ it tends to a fixed point (which position in the phase space is determined by the initial conditions and the certain layout of switching parameters).

In networks defined on the inhomogeneous directed graphs depleted in cycles, no oscillations arise in the system even if the negative interactions in between genes present therein in abundance ($\eta_c=0$). For such networks, the bidirectional edges (if occur) influence on the dynamics essentially. In particular, if a number of edges in the "maximal" graph is  bidirectional, oscillations can arise and persist in the system at any low rate of negative interactions between genes ($\eta_c=1$).

Local dynamics observed in the inhomogeneous scalable regulatory networks is less sensitive to the choice of initial conditions. The scale free networks demonstrate their high error tolerance.

\end{abstract}
\vspace{0.5cm}

\textbf{Keywords:} \textit{Gene regulatory networks, mathematical modelling, mathematical biology.}
\vspace{0.5cm}


\noindent
\textbf{Properties of dynamical networks attract the close attention due to a plenty of their practical applications. One class of them is constituted by the growing \textit{evolutionary} networks representing the \textit{long time evolution} of a genome of living spices or such as the World Wide Web where the dynamics and topology evolve synchronously according to the external principles of informational safety and economic efficiency by the adding of new components at a time.}

\textbf{Another class is represented by the \textit{regulatory} networks including a \textit{fixed} number of elements interacting with each other sensitively to the actual position of system in its phase space. In the present paper, we study  the simplest model of such a regulatory network described by a discrete time synchronously updated  array of coupled structural (topological) and dynamical variables defined at each node of a large graph, for both the homogeneous and scalable graphs.}

\textbf{For small gene expression regulatory networks comprising of just a few elements, a direct logical analysis of dynamics is possible, in which their behavior can be understood as to be driven by the positive and negative \textit{feedback circuits} (loops)  \cite{ST}-\cite{TK}.  However, for large gene expression regulatory networks which can consist of thousands of interacting genes, the direct  analysis is very difficult because of their complexity. Being interested in a qualitative description of dynamical behavior exhibited by such large regulatory networks, we turn this problem in a statistical way. We suppose that  any layout of switching parameters governing the sigmoidal type interactions between genes as well as any assignment of gene interaction signs (\textit{stimulation} or \textit{inhibition}) can be possible with some probability. }

\textbf{Another important issue discussed in our paper is the influence of global topology onto the local dynamics observed in the large gene expression regulatory networks. Such an effect emerges in many coupled dynamical systems defined on the graphs with different topological properties, for instance, in the problem of epidemic spreading \cite{M}. Being defined on  the regular arrays and homogeneous random networks, the models of virus spreading predict the existence of a critical spreading rate $\lambda_c>0$ such that the infection spreads and becomes persistent if $\lambda\geq \lambda_c$ and dies out fast when $\lambda<\lambda_c$. Recently, it has been shown in \cite{PS}-\cite{VVB} that a variety of scale free networks is disposed to the spreading and persistence of infections at whatever spreading rate $\lambda>0$ the epidemic agents possess that is compatible with the data from the experimental epidemiology. }

\textbf{The dynamical behavior demonstrated by the discrete time coupled map lattice used in the present paper as the models of interacting genes \cite{LFM} is indeed more complicated than that in the probabilistic susceptible - infected - susceptible model discussed usually in epidemiology \cite{M}.  We show that it is featured by two order parameters, that are, the fraction of positive interactions allowed in between genes,  $0\leq \eta\leq 1,$ and the fraction of bidirectional edges $0\leq \mu\leq 1$ presented in the "maximal" graph.}

\textbf{In the large homogeneous regulatory networks, like those defined on the fully connected graphs or the regular random graphs, in which all edges are considered as bidirectional,  the critical fraction of positive interactions in between  genes at which the oscillations arise and persist in the system is $\eta_c<1$. In the directed inhomogeneous networks, like those defined  on the directed scale free graphs, oscillations die out fast even if the negative interactions between genes present therein in abundance ($\eta_c=0$). However, oscillations arise  at any low rate of negative interactions between genes ($\eta_c=1$) provided the "maximal" graph has a number of bidirectional edges. Bidirectional edges effectively increase the number of circuits presented in the scale free network that is the source of oscillatory behavior. }

\textbf{The proposed approach could help in understanding the behavior of large gene expression regulatory networks for lack of actual empirical data concerning the large genomes of living organisms. }


\section{Introduction}
\noindent

An impressive progress in understanding of specific biological regulatory mechanisms which play an important role in the way numerous molecular components interact have been made recently. 
This would be a key to control the development and physiology of a whole living organism. 

Nevertheless, the behavior of large gene expression regulatory networks is still far from being understood mainly because of two reasons: first, up to now, a little is known of the global structure of genetic networks. There is still no a common opinion on  whether they are organized hierarchically (say, as a highly inhomogeneous scalable metabolic network as reported in \cite{JTAOB,JMOB}) or contains a plenty of cross interactions, might be of  quite irregular structure, organized in a form of connected sets of small sub-networks.  Second, the relations between the global structure of networks  and their local dynamical properties are also still unclear.

A useful approach to the regulatory networks comprising of just a few elements consists of modelling their interactions by Boolean equations. In this context, \textit{feedback circuits} (the circular sequences of interactions) have been shown to play the key dynamical roles: whereas positive circuits are able to generate multistationarity, negative circuits may generate oscillatory behavior \cite{TK}. Genetic networks are represented by the fully connected Boolean networks where each element interacts with all elements including itself. A feedback circuit can be formally defined as a combination of terms of the Jacobian matrix of the system, with indices forming a circular permutation. Flexibility in network design is introduced by the use of Boolean parameters, one associated with each interaction of group of interactions affecting a given element. Within this formalism, a feedback circuit will generate its typical dynamical behavior (either stationary or  oscillating) only for appropriate values of some of its logical parameters, \cite{TR}.

Most often in biology, the interactions between various molecular components can have a definite sign. For any circuit, one can easily check that each element exerts an indirect effect on itself which has the same sign for all elements of the circuit, leading to the definition of the "circuit sign". In fact, this sign only depends on the parity of the number of negative interactions involved in the circuit: if this number is even, then the circuit is positive; if this number is odd, then the circuit is negative, \cite{T}.

What makes these concepts important is that specific biological and dynamical properties can be associated with each of theses two classes of feedback circuits. The relation between the presence of positive feedback loops and the occurrence of multiple states of gene expression has been at a focus of investigations for several years (see \cite{TSRT} and references therein). In particular, it has been proven that the presence of positive loop(s) is a necessary condition for multistationarity,  and the negative circuits (with two or more elements) are required for the stable periodicity of behavior, \cite{Th}. Biologically, this means that positive circuits are required for differentiative decisions and negative circuits are for the homeostasis, \cite{PMO}-\cite{TSRT}.

Nevertheless, for the large regulatory networks comprising of many hundreds or even thousands of elements, the detailed logical analysis of possible feedback circuits seems  to be impossible now, since the effect that the element can put on itself indirectly, in large regulatory networks, may follow along many different pathways (indeed, if the interaction network is enough dense) engaging probably a plenty of distinct  feedback structures at once. Furthermore, numerical observations over the various large  discrete time regulatory  networks convinced us that the "functionable" circuits and the rest "passive"  elements  are tightly related to each other in a way that they  play the role of "stabilizers" for the active circuits: a dislocation made to an element of the inactive circuit may stampede a change to some "functionable" circuits and even cause they dissolve.

In the present paper, we focus our attention on the large gene expression regulatory networks defined on some "maximal" graph $ \mathbb{G}$. To get a qualitative understanding of their dynamical behavior, we consider a statistical  ensemble of such regulatory networks, in which the thresholds assigned to 
each pairwise gene interaction and its sign are considered as the discrete random variables taking their values in accordance to some statistical laws. Let us note that at present the values of regulatory parameters driving the behavior of actual gene expression regulatory networks are mostly unknown, so that it would be interesting to test the \textit{sensitivity} of local dynamics observed in large regulatory networks to the random change of switching parameters, for the different types of such networks.

Starting from the fixed initial conditions, in  large gene expression regulatory networks, we shuffled these switching parameters randomly and, otherwise, randomized the initial conditions for a fixed layout of thresholds and interaction 
signs. The  long time dynamical behavior observed in such
a statistical ensemble of large regulatory networks depends essentially upon the topology of underlying "maximal" graph including all possible pairwise interactions in between genes of the given set. Short transient processes arisen in such systems at the onset of simulations conclude into the statistically stable behavior, that is, either a stationary configuration or the multi-periodic oscillations occupying up to a half of system size, in the homogeneous regulatory networks having a plenty of negative interactions.

In contrast to the spreading of chaotic state over the regular and random arrays of piecewise linear and logistic discrete time coupled maps studied extensively in the last  decade \cite{CM1}-\cite{VSBC}, oscillations arisen in the discrete time regulatory networks do not propagate over the whole system and are bounded merely to the oscillating domains. Lack of negative interactions  and directed cycles in the networks brings it into one of fixed points which position in the phase space of system depends upon the certain choice of initial conditions and the layouts of switching parameters. The structure of active subgraph of $\mathbb{G}$ in the homogeneous regulatory networks settled at a fixed point, resembles that one of Erd\"{o}s-R\'{e}nyi's random graphs, \cite{ER}.

The plan of paper is following, in Sec.2, we define the models of large synchronized regulatory networks defined on the both homogeneous and inhomogeneous "maximal" graphs where the switching parameters are shuffled randomly. In Sec.3, we present the results of numerical simulations on large regulatory networks. In Sec.~4, we introduce the mean field approach to the stochastic ensembles of large discrete time regulatory gene expression networks  with randomly shuffled thresholds of  type interactions between genes and their interaction signs. Then we conclude in the last section.

\section{Definition of models}
\noindent

We define the regulatory gene expression network $\mathcal{R}$  on the directed graph  $ \mathbb{G}$ with the set of nodes $\mathcal{V}=\{\mathrm{v}_1,\ldots \mathrm{v}_N\}$ connected by the edges $\mathrm{e}_{ij}\in \mathcal{E}\subseteq \mathcal{V} \times \mathcal{V}$ representing the action of gene $i$ onto gene $j$ (the self-action of genes is possible and corresponds to the loops in $\mathbb{G}$). We call $ \mathbb{G}$ as a \textit{maximal} graph since it contains \textit{all} possible interactions in between  genes of the given set.

The regulatory principle of gene expression networks is that the protein  synthesis rate of a gene is affected (either \textit{stimulated} or \textit{inhibited}) by the proteins synthesized by other genes provided their instantaneous concentrations are below (or over) some threshold values. We assign the positive sign $S_{ij}=+1$ to an interaction if $i$ stimulates the synthesis of protein  in $j$, and  the negative sign $S_{ij}=-1$ otherwise. Indeed, in the real genome, the rate of protein synthesis varies from pair to pair of interacting genes, however, for a simplicity, in the present paper, we suppose that all interactions between genes are of equal \textit{effectiveness}, so that all edges presented in $\mathbb{G}$ have the equal weight.

The maximal graph  $\mathbb{G}$ is specified by its \textit{adjacency} matrix $\mathrm{A}^{0}$ such that $A^0_{ij}=1$ if $\mathrm{e}_{ij}\in  \mathbb{G},$  and $A^0_{ij}=0$  otherwise.
Since the effect of interaction between two genes can be negligible at time $t>0$ (that is, this interaction is \textit{switched off} at time $t$), one can define an \textit{active subgraph}, $\mathbb{G}^t \subseteq \mathbb{G}$ including all interactions \textit{efficient} at time $t>0$ specified by the instantaneous adjacency matrix $ \mathrm{A}^t$.

Following \cite{LFM}, in the present paper,  we consider the synchronized model of gene-gene interactions, that is, time is discrete and the state of system at time $t+1$ is a function of its state at time $t,$ in the form of a coupled map lattice. For each gene $i,$ we define two variables: $x^t_i \in [0,1]$, the relative \textit{concentration} of protein expressing it at time $t>0$, and $y^t_i\in[0,1],$  the \textit{exertion} of gene $i$, that is, an effective action of other genes onto $i$ at time $t$
which depends upon the their relative concentrations $x_j^t$ at  same time $t$.

In the homogeneous regulatory networks (like those defined on the complete graphs $\mathbb{G}(N)$ or on the random regular graphs $\mathbb{G}(N,r)$ with the fixed connectivity $r>3$, \cite{RGBook}) the exertion $y^t_i$ can be defined as the fraction of active incoming edges (i.e., the actions of other genes onto $i$) at time $t>0$,
\begin{equation}\label{ex00}
y^t_i= \left.{\sum_{j=1}^N A^t_{ji}}\right/{\sum_{j=1}^N A^0_{ji}}.
\end{equation}
The elements of $\mathrm{A}^t$ are updated synchronously at each time step, in accordance to the current values of $\mathbf{x}^t\in [0,1]^N,$  
\begin{equation}
\label{Th}
A^t_{ij}=A^0_{ij}{\ }\theta\left(S_{ij}(x^t_i-T_{ij})\right),
\end{equation} 
in which the step function $\theta(z)=1,$ for $z>0,$ and $\theta(z)=0,$ for $z\leq 0$, ${\ }T_{ij}\in [0,1]$  is the threshold value for the action of $i$ onto $j$, and $S_{ij}=\pm 1$ is the sign of interaction. In accordance to (\ref{Th}), the interaction $i\to j$ is \textit{active} at time $t$ if either $x^t_i> T_{ij}$ for $S_{ij}=+1$ or $x_i^t< T_{ij}$ for $S_{ij}=-1.$ In the case of $x^t_i=T_{ij}$, we suppose that the action is active provided that  $S_{ij}=+1.$

Then, the discrete time synchronous coupling, 
\begin{equation}
\label{x} 
\mathbf{x}^{t+1}=a \mathbf{x}^t+(1-a)\mathbf{y}^t, \quad 0\leq a<1,
\end{equation} 
generates the flow $\phi^t$ in the phase space 
$[0,1]^N\times[0,1]^N$ transforming the initial point $(\mathbf{x}^0,\mathbf{y}^0)$ into $(\mathbf{x}^t,\mathbf{y}^t)$ for some $t>0$. The parameter $0\leq a<1$ is a positive  constant
expressing the decay rate of protein concentrations, the second term in (\ref{x}) describes the  rates of protein synthesis.
The protein decay rate $a$ determines a time scale in the system, $t\to t(1-a),$ and does not affect its stable asymptotic  behavior. For the first time, such a  model  has been discussed in  \cite{LFM} where the dynamics of several low dimensional examples of genetic expression networks have been considered.

It is important to note that the precise values of switching parameters $T_{ij}$ and $S_{ij}$ are  at present  unknown for the most of pairwise interactions of genes in the genomes of living organisms. Being interested in a qualitative  behavior exhibiting by the large gene expression regulatory  networks, we consider these switching parameters as the discrete random variables, i.e. we suppose that \textit{any}  layout of them is possible with some nonzero probability, and study the statistical behavior of the flow  $\phi^t$ over the ensemble of such networks. Namely,  we suppose that the threshold assigned to a pairwise interaction of genes can take one of $n\leq N^2$ possible values,  $0\leq T_1< \ldots < T_n\leq 1,$ distributed over the unit interval. We  also suppose that the interaction sign $S_{ij}$ takes the value $+1$ with some probability $0\leq \eta\leq 1,$ and $S_{ij}=-1$ with the probability $(1-\eta).$
In the numerical simulations, we checked out $500$ different \textit{random} strings of initial conditions  $ \mathbf{x}^0$ for each of $500$ random layouts of $T_{ij}$ and $S_{ij}$.

Dynamical behavior exhibited by the model with the random layouts of switching parameters depends very much on the topology of  maximal graph. Many \textit{real-world} networks and gene expression regulatory networks, in particular, can be represented by the inhomogeneous graphs of very complex, may be irregular structure  having a  large number of nodes. They can be considered as the \textit{quasi}-random graphs characterized by some "statistical" properties (for instance, by the probability distributions of  degrees of their nodes) \cite{N2001}.

It is understood \cite{T} that the formation of feedback circuits driving the dynamical behavior in the regulatory networks is closely related to the occurrence of cycles in the underlying "maximal" graphs. In particular, it restricts the application of some random processes (algorithms) used for the generation of scale free graphs appropriate for such simulations. For instance, it is obvious that the popular preferential attachment algorithm proposed in \cite{BA}, in which vertices are added to the graph one at a time and connected to a fixed number of earlier vertices, selected with probabilities proportional to their degrees, is not suitable for the generation of such a "maximal" graph, for the gene expression regulatory networks, since  it creates a tree like structure with no cycles. In the present paper, we use another graph generating algorithm (details are given in the Appendix), proposed in \cite{VB} and then used in \cite{VVB} for the study of epidemic spreading over the scale free networks. It creates a directed graph with a bounded number of cycles for which both the incoming and outgoing degree statistics are scale free.

In the gene expression regulatory networks defined on the inhomogeneous graphs, the incoming and outgoing node degrees may differ (like in the directed random \textit{scale free} graphs), so that the definition (\ref{ex00}) of exertions $y^t_i$ has to be improved, since, first, some nodes in such graphs may have only the outgoing edges (\textit{regulating} genes), $K_i^{-}=\sum_{j=1}^N A^0_{ij}>0,$ while $K_i^{+}=\sum_{j=1}^N A^0_{ji}=0$. Second, the stability of numerical scheme  which ensures that $x_i^t\in [0,1]$ for any $t>0$ requires  the factor $(K^{+}_i+K^{-}_i)^{-1}$ before the protein synthesis term in (\ref{x}).

In the case of directed scale free "maximal" graph, the behavior of the minimally required extension of  model (\ref{ex00}), 
\begin{equation}\label{exx}
  y_i^t=\frac{\sum_{j=1}^N A^t_{ji}}{\sum_{j=1}^N A^0_{ji}+A^0_{ij}},
\end{equation}
is not very interesting from the dynamical point of view, since
all nodes with $\sum_{j=1}^N A^t_{ji}\leq K_i^{+}=0$ are decoupled from the rest of network, and there are also many nodes (hubs or regulating genes) with an excessive number of outgoing edges, $K_i^{-}\gg 1,$ for which $y^t_i\ll 1$, so that they are also \textit{effectively} decoupled from other genes in the network. 

Given a scale free random graph with the probability degree distributions $P(K^{\pm})\sim (K^{\pm})^{-\gamma}$ where $\gamma>1,$ then the probability to observe the exertion $y^t>1/2$ in (\ref{exx}) scales with $K^{-}$ like $P(y^t>1/2)\leq (K^{-})^{-2\gamma+1}\sum_{m=0}^\infty (-1)^m(m+\gamma)^{-1}\ll 1,$ for $K^{-}>1,$ and decreases rapidly with $\gamma$. As a result, the most of protein concentrations in the model (\ref{x}) with the exertion (\ref{exx}) defined on the directed scale free graphs are driven by their own decays and got fixed fast close to $\mathbf{x}_{*}=0$. In particulary, one cannot observe oscillations (as a limiting stable behavior) in the regulatory gene expression networks (\ref{Th}-\ref{exx}) defined on such graphs with $\gamma>1$, for any layout of switching parameters and for any assignment of  interaction signs.

The occurrence of \textit{bidirectional} edges (when both genes can act onto each other simultaneously) in the highly inhomogeneous scalable graphs can change dramatically the dynamical behavior of large regulatory networks defined on them.  
To quantify the value of bidirectional edges in the given "maximal" graph, we introduce the parameter $0\leq \mu\leq 1$, that is, the fraction of such edges among all edges of $\mathbb{G}.$

We have performed the numerical simulations for both the homogeneous graphs (the complete graphs and the "enough dense" regular random graphs comprising of $10^3$ nodes) assuming all edges in them to be bidirectional, with no loops, and the highly inhomogeneous scalable graphs, the directed scale free graphs $\mathbb{G}(10^3,2.2)$  such that both probabilities that a node has precisely $K_{+}$ incoming edges and $K_{-}$ outgoing edges follow the power law,  $P(K_{\pm})\sim K_{\pm}^{-2.2}$. This graph has been reported in \cite{JTAOB} as being typical for  the metabolic networks of many living organisms. In the latter case,
we have varied the fraction of bidirectional edges in the whole interval $0<\mu<1$.

\section{Numerical results}
\noindent

We monitor the system approaching a statistically stable regime by tracking out its "velocities" in the phase space, $\mathbf{v}^t=\mathbf{x}^{t+1}-\mathbf{x}^t$ (the rate of protein synthesis) and $u^t=\sum_{ij}\left|A^{t+1}_{ij}- A^t_{ij}\right|$ $/N^2$ (the rate of gene exertions) counting the number of edges triggered between $ \mathbb{G}^t$ and $ \mathbb{G} \setminus \mathbb{G}^t$ at time $t$.       
While studying the model of large regulatory networks (\ref{Th}-\ref{x}) with the random layouts of switching parameters defined  on both homogeneous and inhomogeneous graphs  consisting of $10^3$ nodes, we have varied the  number $n$ of distinct threshold values, $0\leq T_1<\ldots <T_n\leq 1$, from several tens to several hundreds changing by this way the coarse graining of  phase space.  We choose the threshold values $\{T_1,\ldots, T_n\}$ uniformly distributed (u.d.) over the interval $[0,1]$. In fact, in the actual simulations,  we have assumed the "microcanonical" distribution of thresholds characterized by the probability density function $g(x)=n^{-1}\sum_{k=1}^n \delta (x-k/n),$ where $\delta(x)$ is the Dirac function, so that each "event" $T_k=k/n$ had the same statistical weight, $n^{-1}$.

To display the dependence of dynamical behavior on the random 
initial conditions or the random layouts of switching parameters
or both, we have presented the collected statistical data either in the form of density plots or the box-plots where a central line of each box shows the median (i.e., the middle value 
collected over $500$ random initial conditions or random layouts), a lower line of a box  shows the first quartile,  and an upper line of a box shows the third quartile. Half of the difference between the third quartile and the first quartile (the semi-interquartile range or the quartile deviation) is a measure of the dispersion of the data. Two lines extending from the central box of maximal length 3/2 the interquartile range.

We commence the analysis with the networks defined on the undirected homogeneous graphs: the complete graph and the regular random graph in which the connectivity of each node is fixed at some  $K\geq 3$ to ensure the existence of many Hamilton cycles traversing all nodes in the graph. All nodes of such graphs have the equal connectivities $K$ and each node is connected to any other with some probability $p>0$ (for the fully connected graph, $K=N-1$ and $p=1$). The behavior of the model (\ref{ex00}-\ref{x}) defined on them looks essentially similar: after a considerably short transient process, the system settles into a statistically stable dynamical  regime which depends very much upon the fraction of negative interactions allowed in between genes and quantified by $1-\eta$.

\begin{figure}[ht]
 \noindent
 \begin{minipage}[b]{0.36\linewidth}
 \begin{flushleft} 
\begin{tabular}{rlrl}
 a. &\epsfig{file=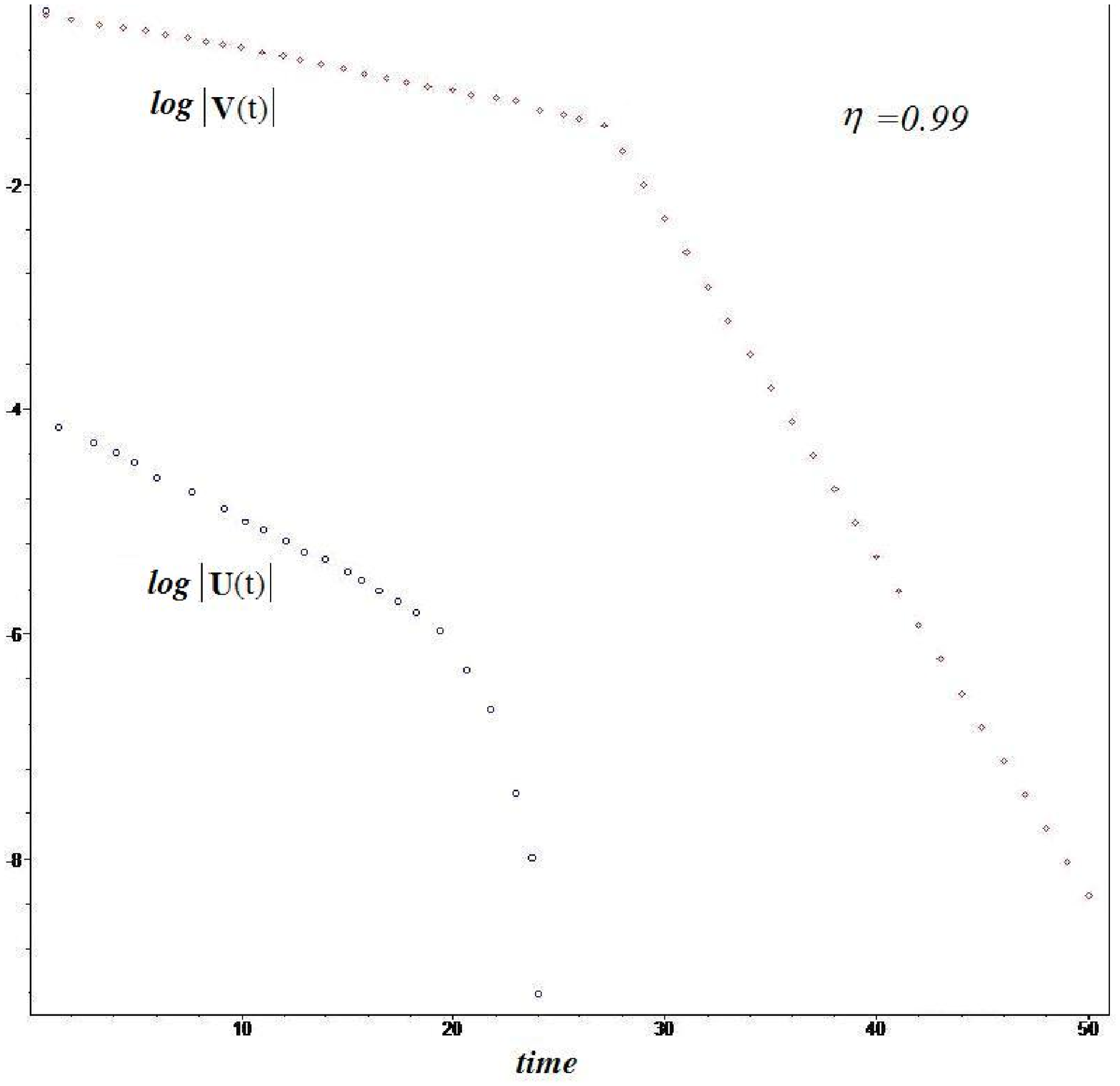, angle= 0,width=8.0cm} &
 b.&  \epsfig{file=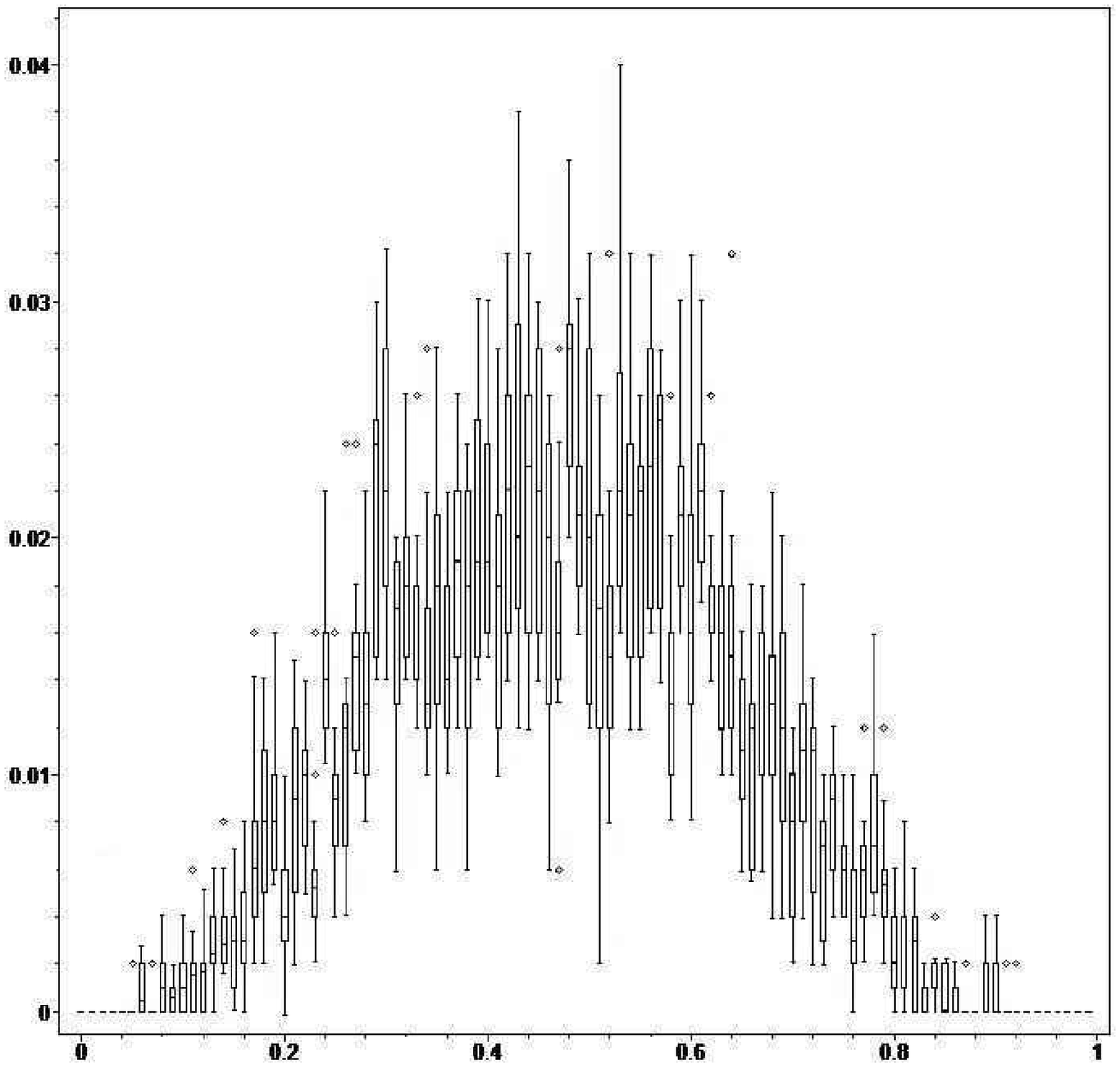, angle= 0,width=8.0cm}
 \end{tabular}
 \end{flushleft}
\end{minipage}
\caption{\label{fig1} The behavior of model (\ref{Th}-\ref{x}) for a given random layout of switching parameters defined on the fully connected graph $\mathbb{G}(10^3)$ in the vicinities of its fixed points. 
a). The exponentially fast approach of the rate of synthesis $|\mathbf{v}^t|=|\mathbf{x}^{t+1}-\mathbf{x}^t|$ and 
$u^t=\sum_{ij}\left|A^{t+1}_{ij}- A^t_{ij}\right|/N^2$
to zero for the model with a fixed configuration of $50$ thresholds uniformly distributed (u.d.) over the unit interval
and the fixed assignment of $S=-1$ to $1\%$ of interactions, $\eta = 0.99$, the protein decay rate $a=0.7.$ 
b). The occupancy number $p_k$ of the consequent intervals $\Delta_k=[T_{k-1},T_k]$ at the different fixed points of  model with a given configuration of $100$ thresholds u.d. over the unit interval, $\eta=0.95$ ($5\%$ of negative interactions), $a=0.7$. The strong fluctuations shown by the boxes reveal the dependence of the occupancy numbers upon the certain choice of initial conditions; data has been collected over $500$ random initial strings  $\mathbf{x}^0,$ for a fixed layout of switching parameters.}
\end{figure}

Provided the system has just a few negative interactions in the network ($\eta \to 1$), it approaches one of the  fixed points exponentially fast, for any random initial condition $\mathbf{x}^0$ (see Fig.~\ref{fig1}.a). Herewith, the rate of synthesis of proteins ($|\mathbf{v}|^t$) shows a crossover between two consequent phases of transient regime (see Fig.~\ref{fig1}.a): before and after the moment when the structure of active subgraph $\mathbb{G}^t$ gets fixed. The exponential decay rate of transients is independent on neither the choice of initial conditions nor the certain layout of switching parameters, but the stationary asymptotic configuration $\mathbf{x}_{*}$ depends upon both.

For a given random layout of switching parameters, many different stable fixed points can be achieved by the system starting from the different initial conditions $\mathbf{x}^0$ that is an evidence of multistationarity in the model. To get an image of variety of the possible asymptotic stationary states, we have counted the occupancy numbers $p_k=\{\sharp i: T_{k-1}< {x_i}_{*}\leq T_k \}$ of the consequent intervals of phase space, $\Delta_k=[T_{k-1},T_k],$ formed by $100$ threshold values $T_k$ u.d. over the unit interval (see Fig.~\ref{fig1}.b). The boxes shown in Fig.~\ref{fig1}.b present the variations of occupancy numbers $p_k$ for $500$ different random initial conditions $\mathbf{x}^0$ with $5\%$ of negative interactions allowed between genes ($\eta=0.95$), for some fixed layout of switching parameters.

\begin{figure}[ht]
 \noindent
 \begin{minipage}[b]{.36\linewidth}
 \begin{flushleft} 
\begin{tabular}{rlrl}
 a. &\epsfig{file=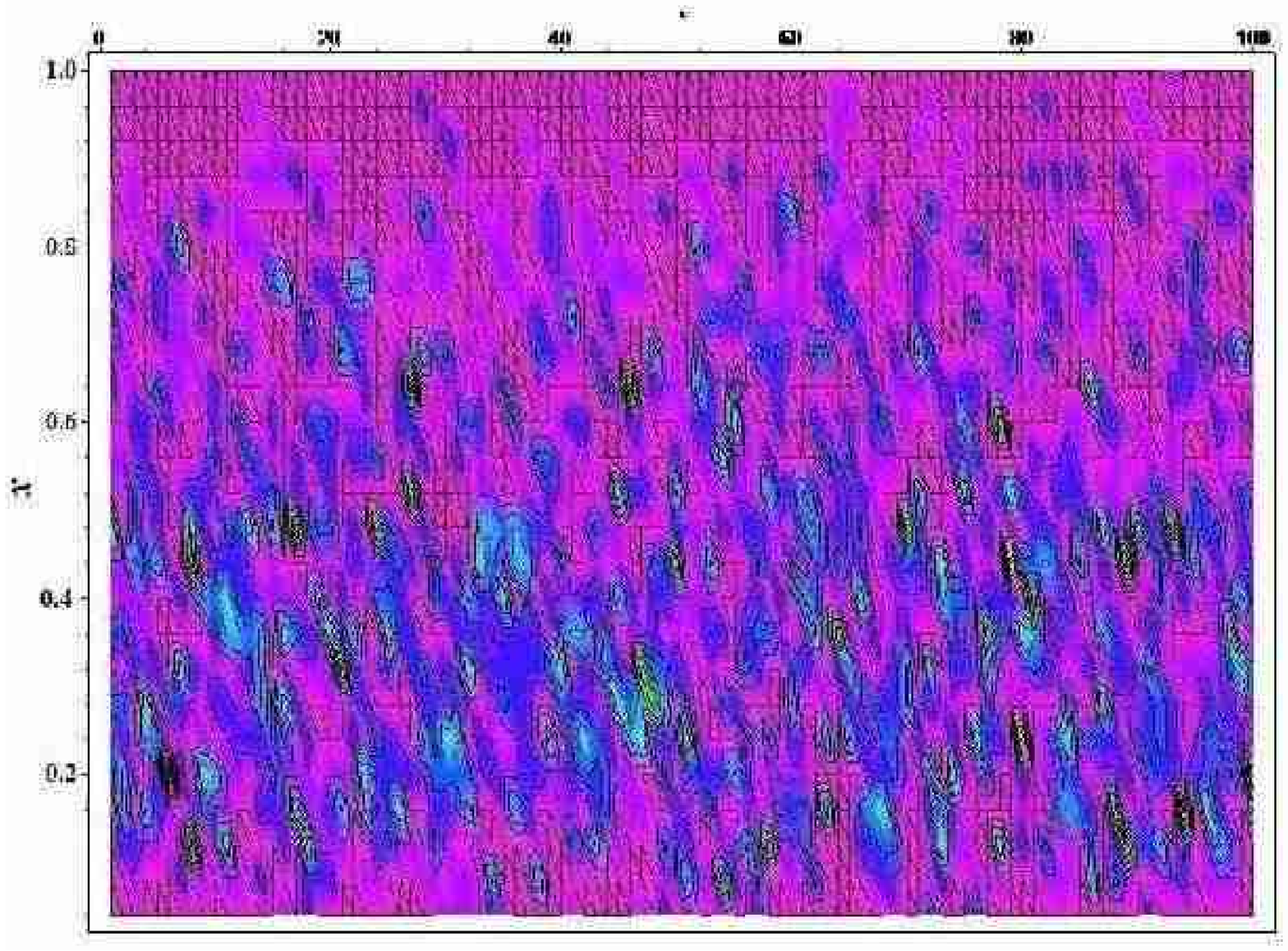, angle= 0,width=8.0cm}&b.&
 \epsfig{file=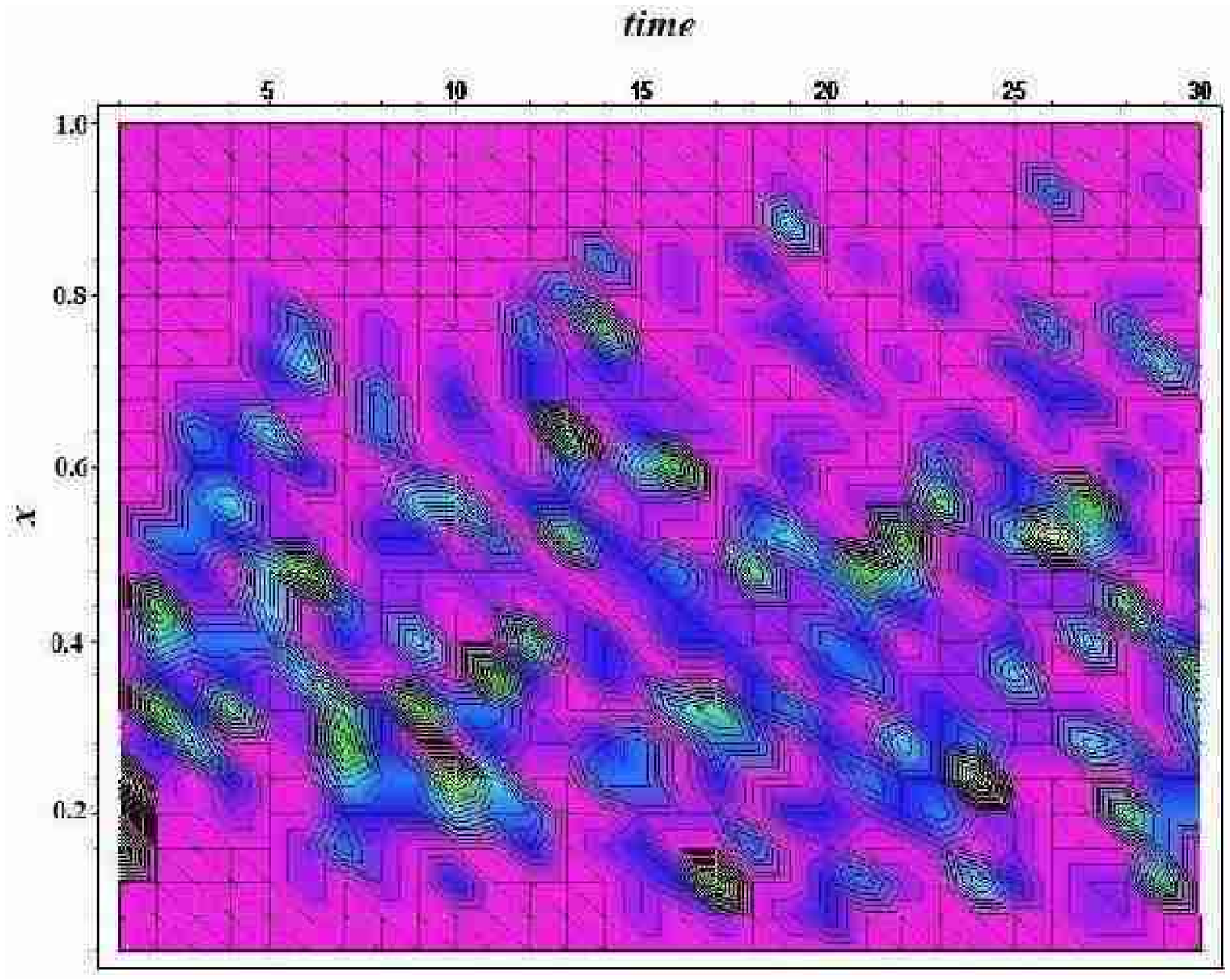, angle= 0,width=7.5cm}
 \end{tabular}
 \end{flushleft}
\end{minipage}
\caption{\label{fig2} The density of stationary configurations  $\mathbf{x}_{*}$ at the fixed points depends strongly upon the certain layouts of switching parameters. 
a). The density plot of empirical distributions of stationary values $x_i{}_{*}$  for $100$ nodes chosen at random among total $10^3$. Stationary configurations have been achieved by the system starting from the given string of initial conditions. $50$ distinct threshold values u.d. over the unit interval have been shuffled ($500$ different layouts) randomly together with the random assignments of $1\%$ of negative interactions ($\eta=0.99$),  $a=0.6$. The patchy structure of graph reveals the multistationarity in the system. Some patches merge manifesting the sensitivity of stationary configurations to the certain layout of switching parameters.
 b). The density plot of empirical distributions of values $x_{77}^t$  vs. time in $30$ consequent time steps (long enough to achieve a fixed point) starting from $x_{77}^0=0.175$ on the 
fully connected graph of $10^3$ nodes.  The random shuffle of switching parameters mixes up the orbits of  deterministic system (\ref{Th}-\ref{x}).
}
\end{figure}

The density of possible  stationary asymptotic configuration $\mathbf{x}_{*}$ depends upon the certain layouts of switching parameters for any certain initial string $\mathbf{x}^0$ that is presented on Fig.~\ref{fig2}. A patchy structure of graph in Fig.~\ref{fig2}.a.  manifests the multistationarity in the system, meanwhile the "clusters" formed by the merged patches show that the limiting stationary configurations $\mathbf{x}_{*}$ are  sensitive to the layout of switching parameters and can move gradually as these parameters shuffle.

Shuffling of switching parameters \textit{mixes up} the orbits of deterministic system intensively, as a result each $x^t_i$ spreads out fast with time over the whole unit interval. In  Fig.~\ref{fig2}.b, we have sketched the density plot of  possible values of $x_{i}^t$ (for $i=77$ in $\mathbb{G}(10^3)$) vs. time in $30$ consequent time steps (long enough to achieve a fixed point) starting from the initial value $x_{77}^0=0.175$.

\begin{figure}[ht]
\noindent
\begin{minipage}[b]{.36\linewidth}
\begin{center} 
\epsfig{file=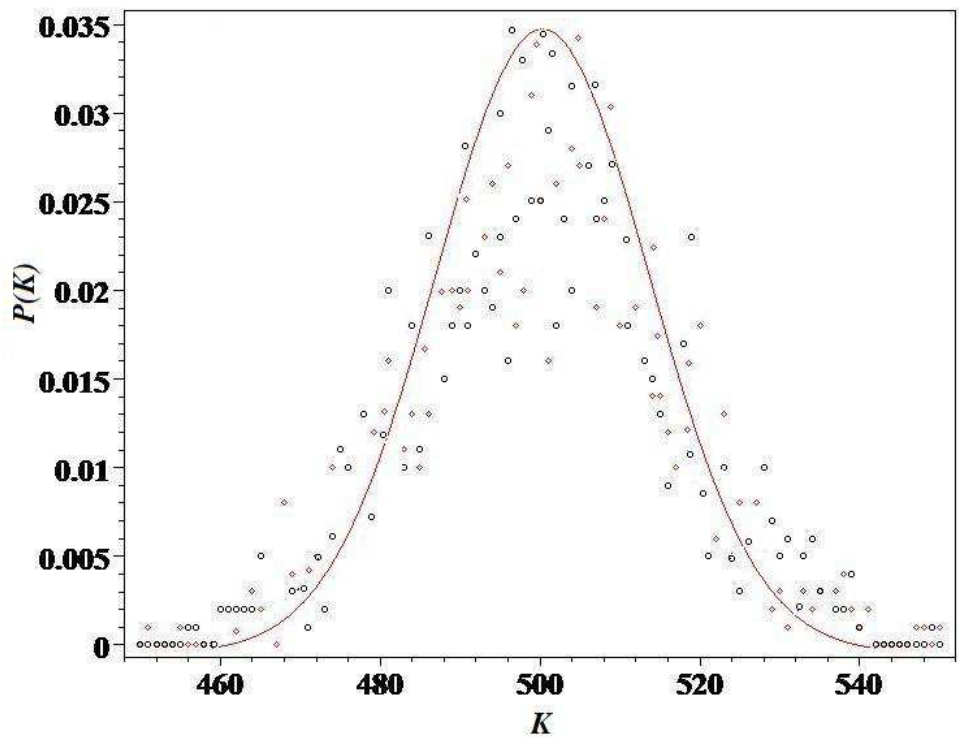, angle= 0,width=13.0cm}
\end{center}
\end{minipage}
\caption{\label{fig3} The probability degree distributions of nodes in the active subgraph formed at a fixed point of the model (\ref{Th}-\ref{x}) defined on the fully connected graph $\mathbb{G}(10^3)$ with $1\%$ of negative interactions allowed   between genes (the circles are for the incoming degrees and the diamonds are for the outgoing degrees of nodes). The solid line is for the Gaussian probability degree distribution which is typical  for the random graphs of Erd\"{o}s and R\'{e}nyi, \cite{ER}.}
\end{figure}

It is worth to mention that at any fixed point the active subgraph $\mathbb{G}_{*}\subset \mathbb{G}$  constitutes a random graph ("half-dense" in comparison with $\mathbb{G}$). In Fig.~\ref{fig3}, we have shown the probability degree distributions (the circles are for the incoming degrees $K_{+},$ and the diamonds are for the outgoing degrees $K_{-}$) for the nodes of active subgraph $\mathbb{G}_{*}$ formed at a fixed point of the model (\ref{ex00}-\ref{x}) defined on the fully connected graph $\mathbb{G}(10^3)$ with $1\%$ of negative interactions allowed   between genes. The solid line on Fig.~\ref{fig3} displays the Gaussian probability degree distribution which is typical for the Erd\"{o}s and R\'{e}nyi random graphs, \cite{ER}.

One can say that in the homogeneous regulatory networks when the positive interaction between genes prevail in the system, its dynamical behavior is dominated by the \textit{positive feedback circuits} responsible for a number of asymptotically stable states (fixed points). Thereat, the strain of negative feedback circuits related to just a few negative interactions  is statistically negligible.

With increasing percentage of negative interactions allowed in the system up to approximately $10\%$, it exhibits a complicated spatiotemporal behavior where the domains of genes with the stationary concentrations of proteins coexist and interleave with those of  periodically oscillating concentrations. In contrast to the spatiotemporal intermittency observed in the synchronously updated discrete time extended dynamical systems defined on the various regular arrays \cite{CM1,Chate1} and on the regular random graphs \cite{VSBC}, the  dynamical state (oscillations)
does not  propagate with time throughout  the regulatory network.

\begin{figure}[ht]
\noindent
 \begin{minipage}[b]{.36\linewidth}
 \begin{flushleft} 
\begin{tabular}{rlrl}
 a. &\epsfig{file=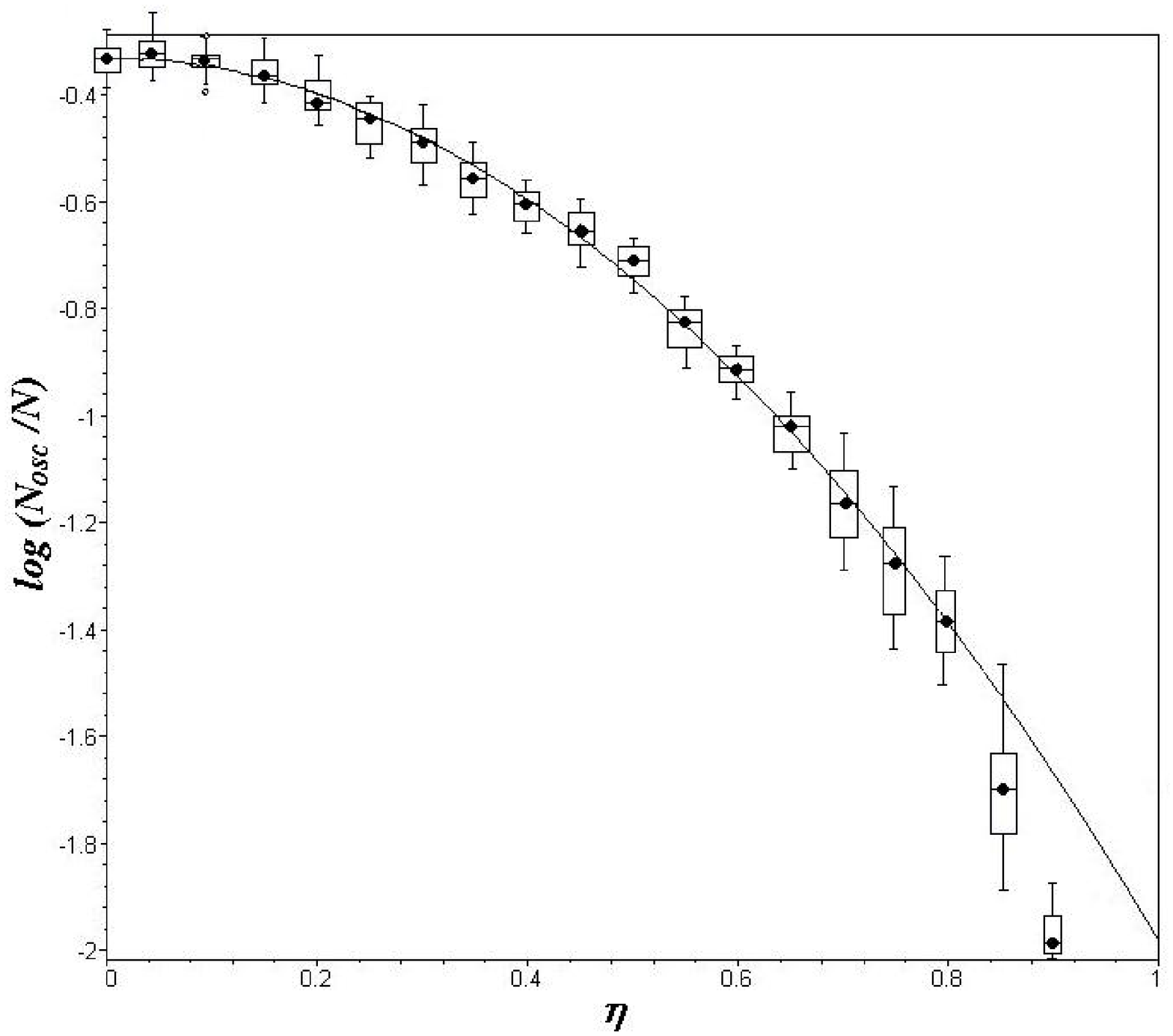, angle= 0,width=7.5cm}&b.&
 \epsfig{file=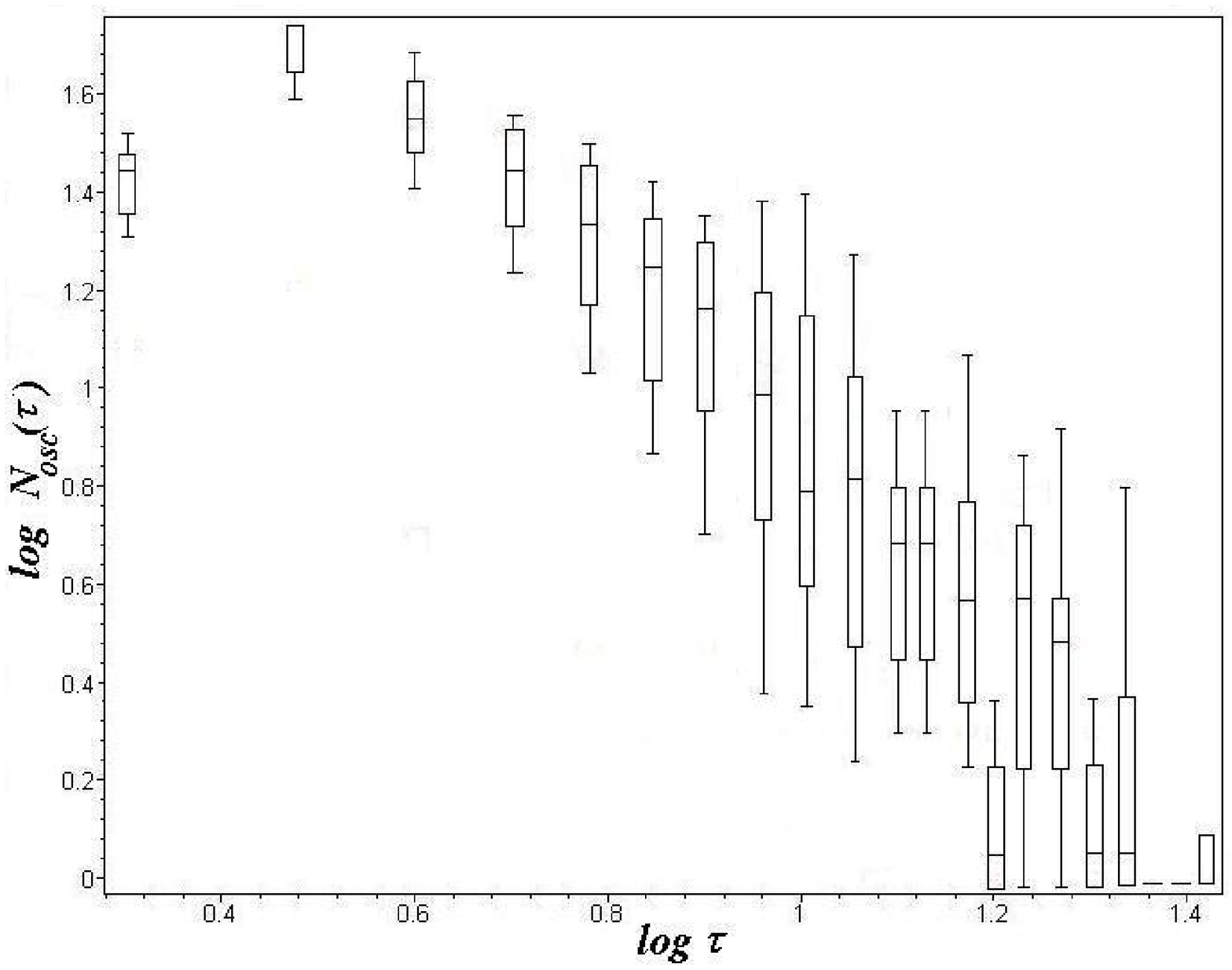, angle= 0,width=8.0cm}
 \end{tabular}
 \end{flushleft}     \end{minipage}
\caption{\label{fig4} a). Decreasing and vanishing of oscillating domains as $\eta$ approaches $\eta_c\simeq 0.92$  in the regulatory network defined on the complete graph $\mathbb{G}(10^3)$. Boxes present the fluctuation of data collected over the ensemble of $500$ different random layouts of switching parameters and initial conditions, $a=0.74$.
The bold points stand for the means. The data are fitted well with the Gaussian curve (the solid line),$\left.N_{ \mathrm{osc}}\right/N \simeq \left.\exp(-\eta^2/2\sigma)\right/\sqrt{2\pi\sigma^2}$ , in which $\sigma\simeq 0.555.$ b). The distribution of  nodes with the oscillating protein concentrations over the periods of oscillations  in the complete graph $\mathbb{G}(10^3)$ at $\eta=0.1$, $a=0.74.$  Boxes present the fluctuation of data collected over the ensemble of $500$ different random layouts of switching parameters and initial conditions.}
\end{figure}

The oscillating domains arisen in the homogeneous regulatory networks are bounded by the genes whose oscillation amplitudes  are insufficient for their protein concentrations to cross the next thresholds. In the large enough homogeneous regulatory networks, a turnover of nodes engaged into the oscillating domains  happens occasionally. The averaged number of nodes joined the oscillating domains at a time increases as the fraction of negative interactions allowed in the model grows up. In Fig.~\ref{fig4}.a, we have displayed the decreasing and vanishing of fractions $N_{\mathrm{osc}}/N$ of oscillating nodes  with $\eta,$ in the model defined on the fully connected graph (with bidirectional edges). Boxes show the fluctuations of these fractions in the ensemble of $500$ different random layouts of switching parameters and random initial conditions. Bold points stand for the means of collected data. 

The solid line in Fig.~\ref{fig4}.a is the Gaussian curve, $\left.N_{ \mathrm{osc}}\right/N \simeq \left.\exp(-\eta^2/2\sigma)\right/\sqrt{2\pi\sigma^2},$ with $\sigma\simeq 0.555$  fitting the data well. The direct logical analysis of low dimensional regulatory  networks (see, for example \cite{TR}) relates  oscillations  to the dynamical patterns generated by the  "functional" negative feedback circuits. A feedback circuit exhibits its typical dynamical behavior  only for appropriate values of some of the logical parameters. The graph sketched on Fig.~\ref{fig4}.a demonstrates that the probability to observe a "functional" negative feedback circuit, in  large homogeneous regulatory networks with the randomly shuffled switching parameters, is close to the normal statistical law with regard to $\eta$.

Fig.~\ref{fig4}.b displays the distribution of nodes changing  their protein concentrations periodically vs. the periods of such changes observed in the large homogeneous regulatory networks. The distribution in Fig.~\ref{fig4}.b counts all such nodes  disregarding for the amplitudes of changes. Each node has been counted in the distribution just once under the minimal period of oscillations it exhibits. As usual, boxes represent the fluctuations of numbers of oscillating genes $N(\tau)$ over the ensemble of $500$ different random layouts of switching parameters and random initial conditions. The distribution has a maximum at $\tau = 3$ independently upon the initial conditions and the layouts of switching parameters.

\begin{figure}[ht]
 \noindent
 \begin{minipage}[b]{.36\linewidth}
 \begin{flushleft} 
\begin{tabular}{rlrl}
 a. &\epsfig{file=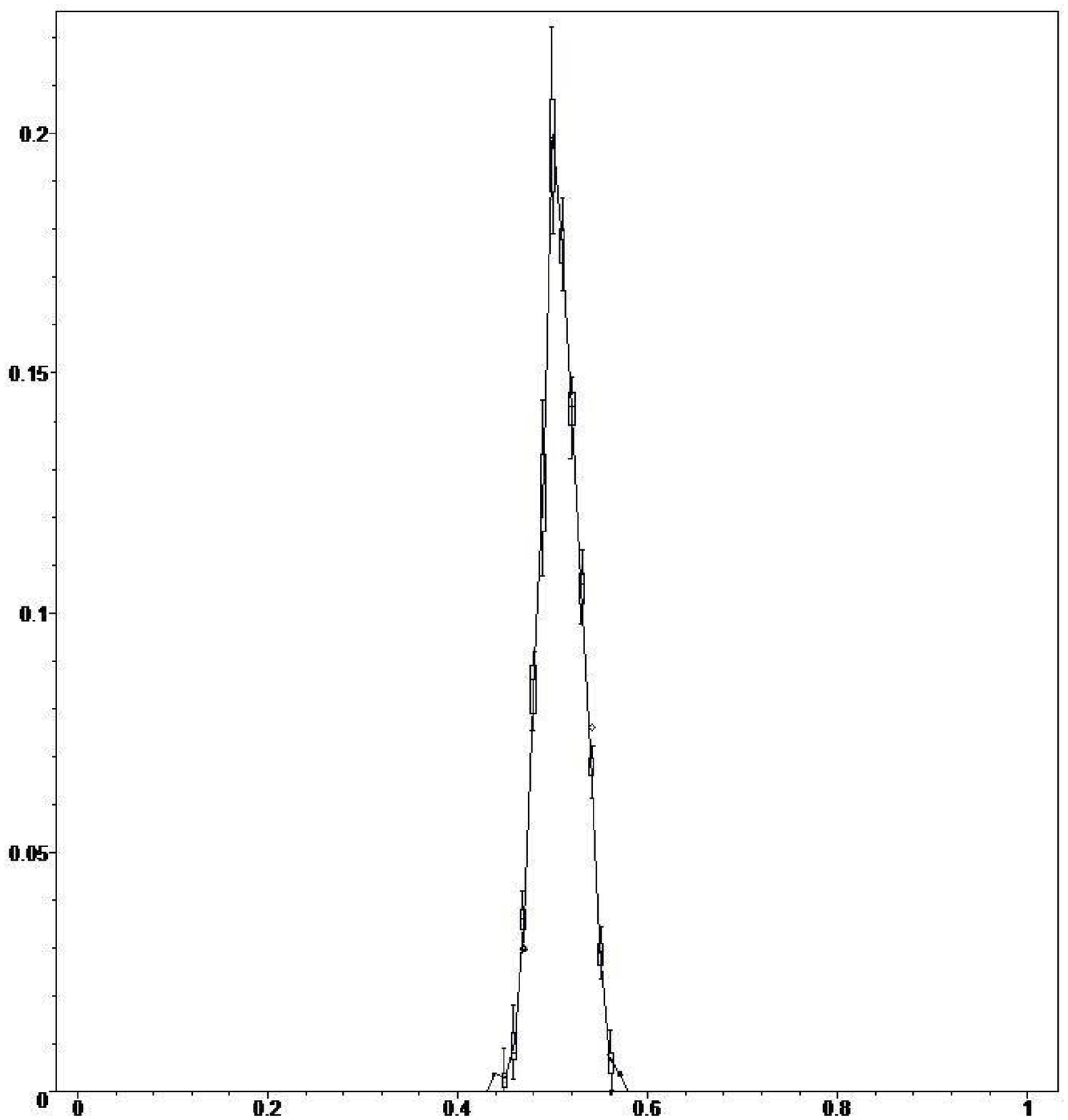, angle= 0,width=7.5cm}&b.&
 \epsfig{file=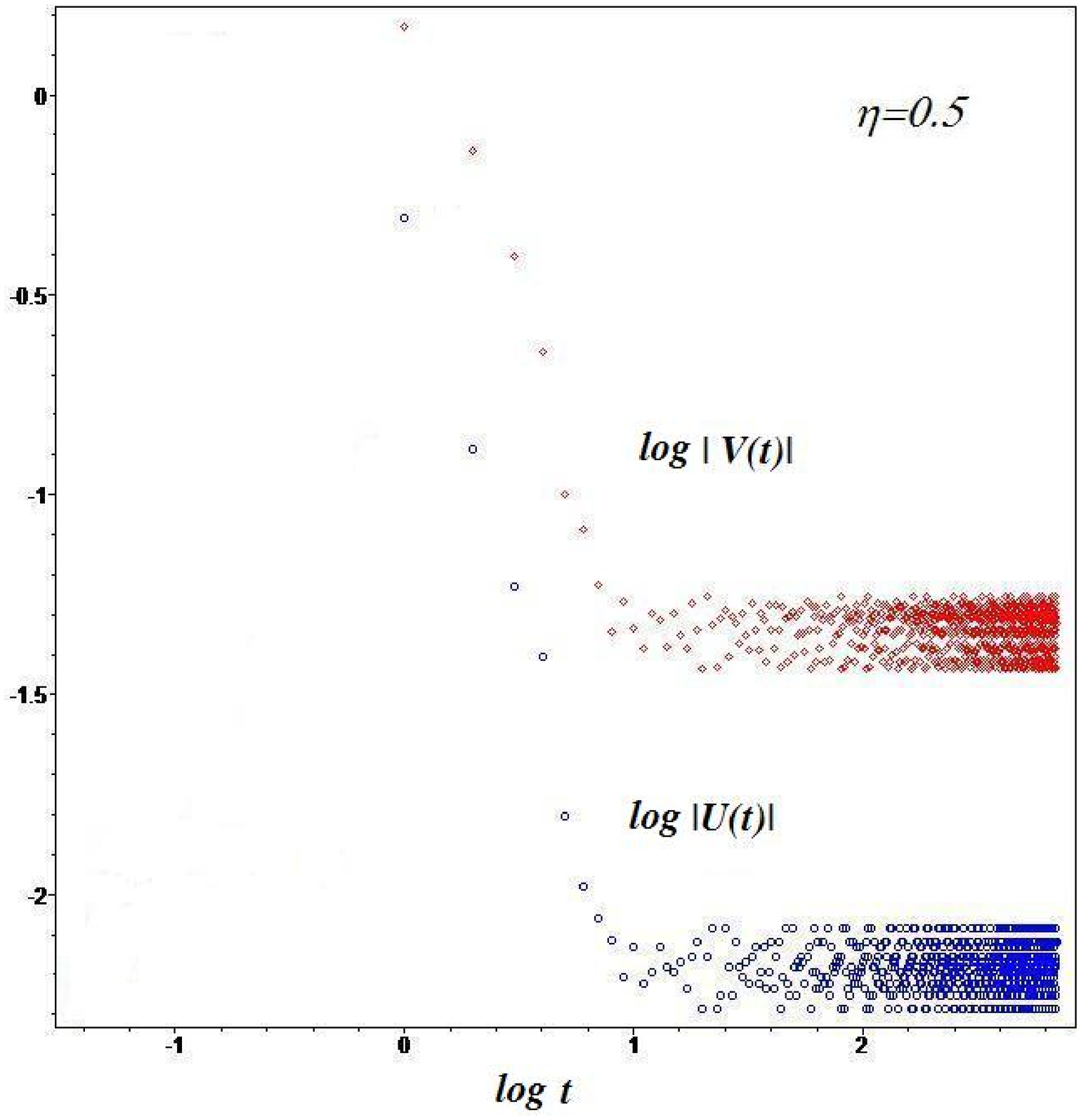, angle= 0,width=8.0cm}
 \end{tabular}
 \end{flushleft}
\end{minipage}
\caption{\label{fig5} a). The occupancy number $p_k$ of the consequent intervals $\Delta_k=[T_{k-1},T_k]$ is  insensitive to the initial conditions for the random layout of switching parameters as $\eta \ll 1$ (computed for the model defined on the complete graph $\mathbb{G}(10^3)$ with $100$ distinct thresholds u.d. over the unit interval, $\eta=0.5$, ${\ }a=0.6$). When oscillations arise, the values $x^t_i$ for the most of nodes (which do not engaged into these oscillations) are synchronized at $ 1/2$. 
b.) When the fraction of negative interactions increases, the transient processes (which decay following a power law as $\eta\to 0$) in the model (\ref{Th}-\ref{x}) defined on the homogeneous graphs conclude in oscillations.}
\end{figure}

Formation  of oscillating domains for $\eta \ll 1$ (when
the negative interactions present in abundance) comes along with the synchronization of the rest of system at $x= 1/2$ (see the profile for the occupancy number $p_k$ in Fig.~\ref{fig5}.a). This synchronization looks essentially insensitive to the initial conditions  (the boxes on the graph are almost imperceptible). It gives  an impression that when the dynamical behavior is obviously driven by the negative feedback circuits, and oscillations of protein concentrations can occupy up to a half of nodes in the network, just a few of them actually change the instantaneous structure of active subgraph $\mathbb{G}^t$. 

 On Fig.~\ref{fig5}.b, we have shown the behavior of the phase space velocities,  $\log |\mathbf{v}^t|$ and $\log |u^t|,$ vs. $\log t$ characterizing the transient processes in the system for the model defined on the fully connected graph  $\mathbb{G}(10^3)$, with $100$ distinct thresholds u.d. over the unit interval, with $\eta=0.5$ ($1/2$ of interactions are negative).  When the negative interactions prevail in the system, these velocities decay much slower than the exponentially fast transients shown in Fig.~\ref{fig1}.a and asymptotically tend to a power law as $\eta \to 0.$ Moreover, they do not extinguish eventually and bring in oscillations in short time.

\begin{figure}[ht]
 \noindent
 \begin{minipage}[b]{.36\linewidth}
 \begin{flushleft} 
\begin{tabular}{rlrl}
 a. &\epsfig{file=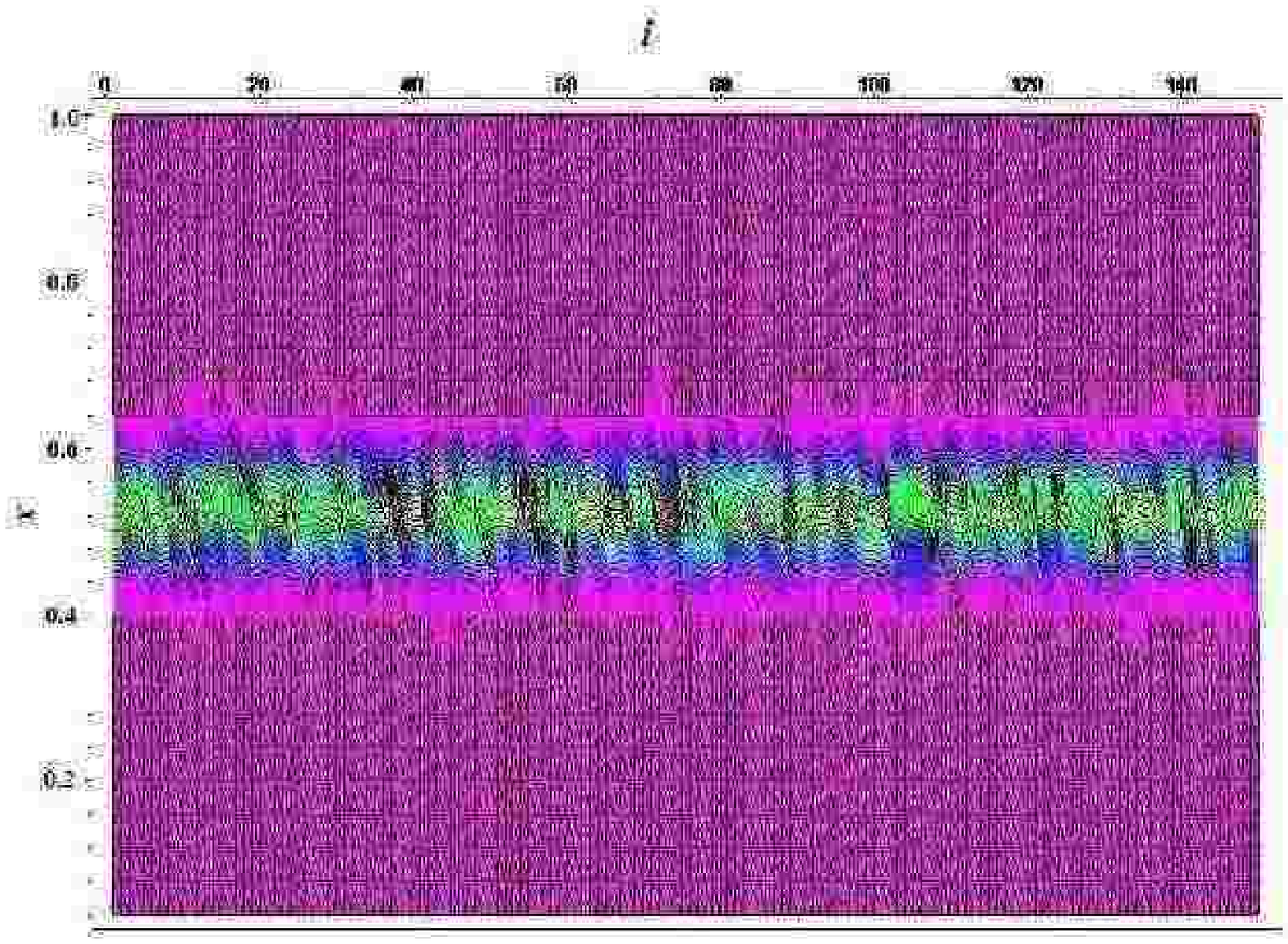, angle= 0,width=8.5cm}&b.&
 \epsfig{file=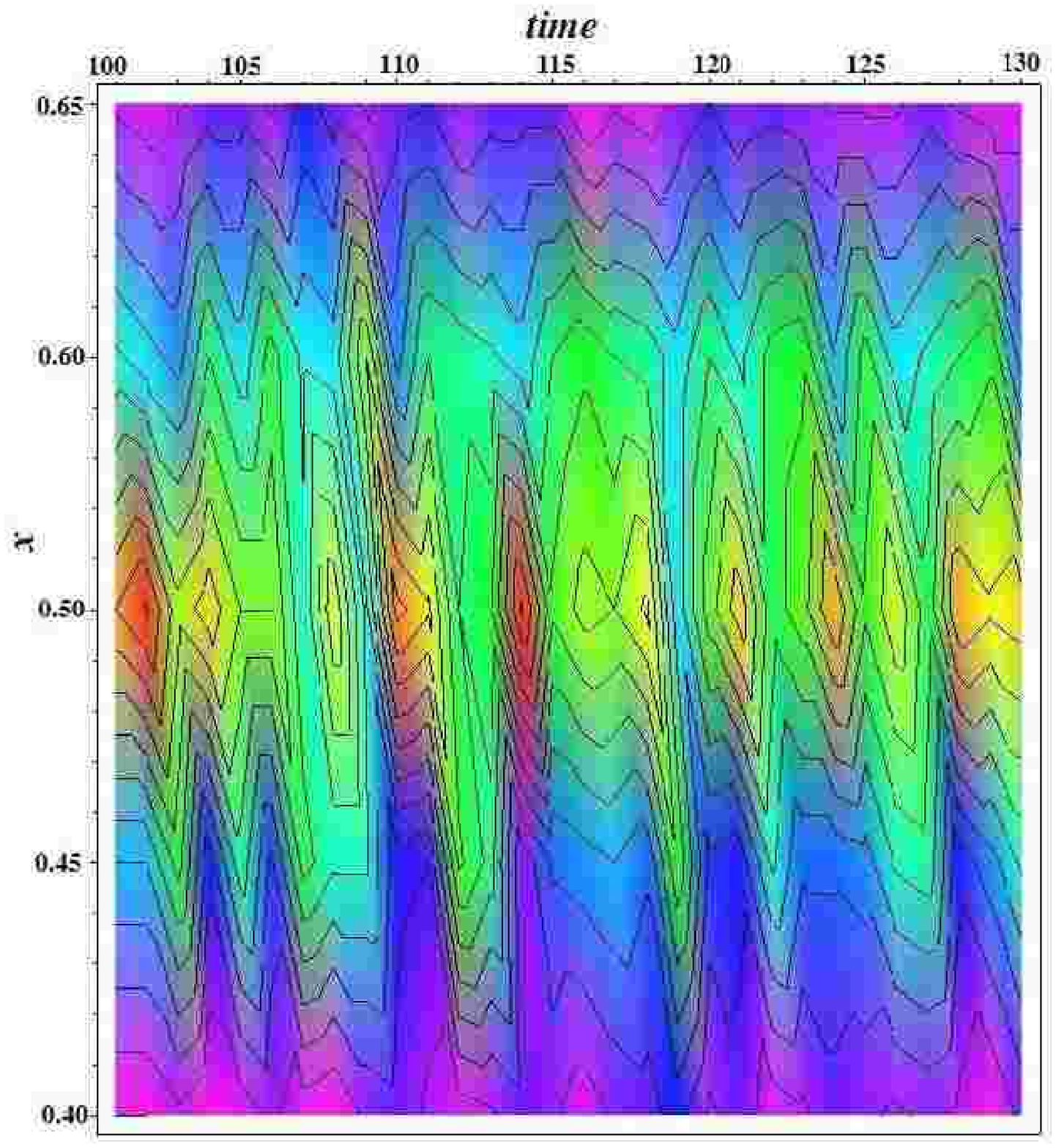, angle= 0,width=7.0cm}
 \end{tabular}
 \end{flushleft}
\end{minipage}
\caption{\label{fig6} Oscillations arise in the system  together with synchronization of the rest of network at $1/2$ is insensitive to the shuffle of thresholds and interaction signs. 
a.) The density plot of empirical distributions of values $x^t_i$  for the first $150$ nodes of the complete graph $\mathbb{G}(10^3)$ in the stable oscillating regime, $\eta=0.5$, ${\ }a=0.6,$ taken over $500$ distinct shuffles of switching parameters. 
b.) The density plot of empirical distributions of values $x_{77}^t$  vs. time in $30$ consequent time steps ($t=100\ldots 130$) taken over the ensemble of  $500$ different random layouts
of $50$ distinct thresholds. }
\end{figure}

Stable patterns of statistical behavior are  insensitive to the random layouts of thresholds and assignments of interaction signs. Starting from some fixed initial string $\mathbf{x}^0$, we have shuffled the switching parameters  in the model defined on  $\mathbb{G}(10^3)$ (with bidirectional edges)  for $\eta=0.5$ and  displayed the data for $x^t_i$ (after the stable behavior had been achieved) for the first $150$ nodes (see Fig.~\ref{fig6}.a).  The time evolution of these density distributions indicating oscillations is illustrated by the beatings shown in  Fig.~\ref{fig6}.b in $30$ consequent time steps taken over the ensemble of $500$ different random layouts of switching parameters.

It is important to note that for any value of $\eta$ 
it is always \textit{less} than $1/2$ of nodes (achieved only if all interactions between genes are negative) that are engaged into oscillations in the statistically stable regime. Herewith, the oscillating protein concentrations $x_i^t$ for the most of nodes are still bounded within their intervals $\Delta_k=[T_{k-1},T_k]$ and do not change the occupancy numbers $p_k$. Direct investigation of active subgraphs in the stable oscillating regimes had shown that for any random initial condition the notable oscillations of protein concentrations resulting in  changes to active subgraph are confined to \textit{less} than $3\%$ of nodes.

\begin{figure}[ht]
 \noindent
 \begin{minipage}[b]{.36\linewidth}
 \begin{flushleft} 
\begin{tabular}{rlrl}
 a. &\epsfig{file=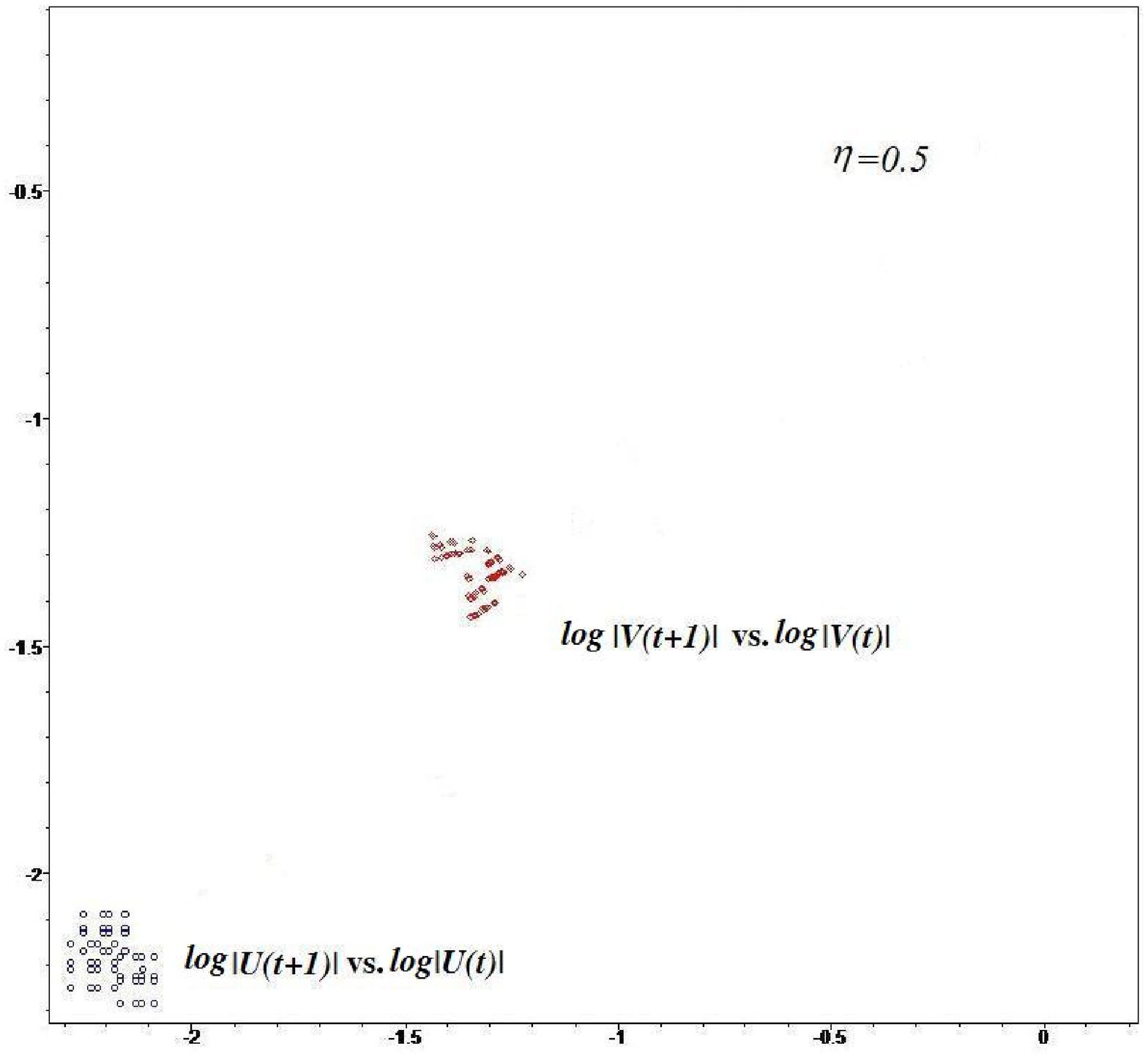, angle= 0,width=7.5cm}&b.&
 \epsfig{file=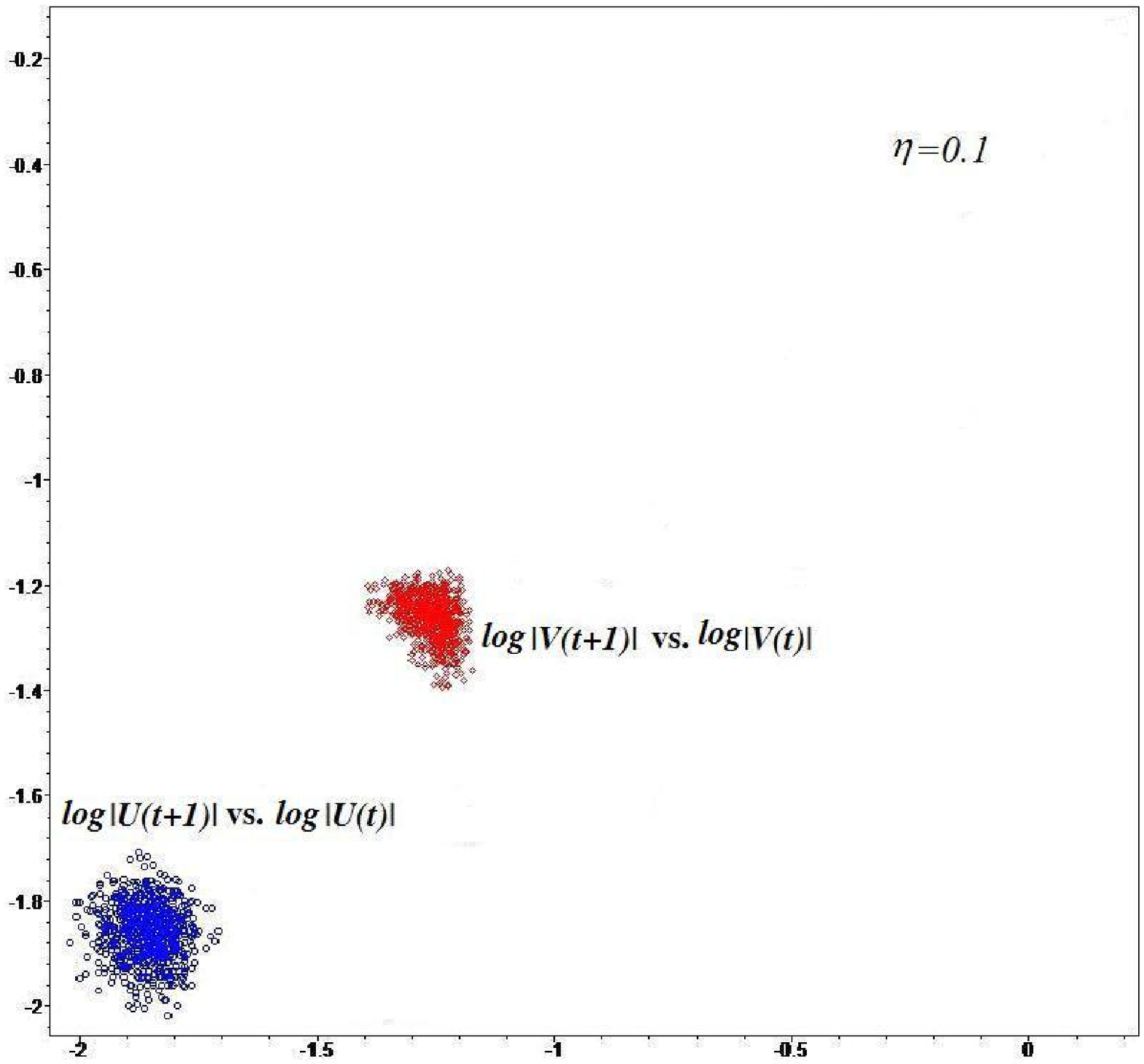, angle= 0,width=7.5cm}
 \end{tabular}
 \end{flushleft}
\end{minipage}
\caption{\label{fig7}
The return maps for the model (\ref{Th}-\ref{x}) defined on $\mathbb{G}(10^3)$ for the fixed layout of $100$ distinct thresholds u.d. over the unit interval and fixed random initial conditions $\mathbf{x}^0$, ${\ }a=0.6,$ reveal the multi-periodicity of oscillations risen in the system, a.) for $\eta=0.5$, b.) for $\eta=0.1$.}
\end{figure}

To illustrate the multi-periodicity and regularity of  oscillations arisen in the model (\ref{Th}-\ref{x}) defined on $\mathbb{G}(10^3)$, we have presented the return maps expressing $\log |\mathbf{v}^{t+1}|$ vs. $\log |\mathbf{v}^{t}|$ and $\log |u^{t+1}|$ vs. $\log |u^{t}|,$ for $\eta=0.5$ in Fig.~\ref{fig7}.a and for $\eta=0.1$ in Fig.~\ref{fig7}.b. Detailed observations of the fine structure of return maps 
in Fig.~\ref{fig7}.a show that when the system  has a parity between negative and positive interactions, it exhibits oscillations of shortest periods, $\tau=2,3,4$. Oscillations of many different periods arise if the negative interactions prevail in the system.

In a conclusion, when the fraction of positive interactions between genes in the regulatory network defined on the homogeneous graphs decreases down to the critical value $\eta_c\simeq 0.92,$ the negative feedback circuits influence the dynamical behavior substantially  that is revealed in persistent oscillations arisen in the system at the end of transient processes.

We have also studied the behavior of  model (\ref{Th}-\ref{x}) 
defined  on the inhomogeneous scale free graph $\mathbb{G}(10^3,2.2)$ with the exertion (\ref{exx}), in which both statistics of incoming and outgoing edges follow the power law $P(K^{\pm})\sim (K^{\pm})^{-2.2}$ shown on Fig.~\ref{fig8}. It has been  generated in accordance to the algorithm given in  Appendix A. 

The behavior demonstrated by the dynamical systems defined on scale free graphs depends essentially on their detailed topology  and can be dramatically different for the graphs generated in accordance to  distinct algorithms even though they enjoy the same probability degree statistics \cite{VVB}. Individual structural properties of scale free graphs is characterized by the certain pair-formation process, in which each vertex $v$ of degree $K$ chooses a set of partners according to a specified $K-$dependent rule describing its "preferential choice". Directed scale free graphs have usually  a few cycles (since  cycles imply a statistical parity between numbers of incoming and outgoing edges and therefore cannot play a key role in the structures of \textit{quasi}-random scalable graphs). 

The feedback circuits formation,  in scale free graphs, relays primarily upon the bidirectional edges connecting  pairs of mutually interacting genes. In Fig.~\ref{fig9}, we have presented a phase diagram displaying the appearance of persistent oscillations in the model (\ref{Th}-\ref{x}) 
defined  on the inhomogeneous scale free graph $\mathbb{G}(10^3,2.2)$ for different fractions of negative interactions $\eta$ and bidirectional edges $\mu.$ It shows that  for the enough fractions of bidirectional edges, in the scalable regulatory network, oscillations can persist even if there is just a few negative interactions allowed in between genes of the network ($\eta_c=1$).

We have studied the statistic of oscillations, in the limiting case of $\mu=1,$ when the scale free graph can be considered as undirected.  Fig.~\ref{fig10}.a displays the decrease  of oscillating domains (with a rate close to linear) in the scalable regulatory network. Fluctuations of data collected over the ensemble of $500$ different random layouts of switching parameters and initial conditions at $a=0.74$ are shown by boxes.  Let us note that Fig.~\ref{fig10}.a reveals the similar effect of global scalable topology of network onto the local dynamics of nodes that has been reported recently in \cite{PS}-\cite{VVB} on the problem of virus spreading in the undirected scalable networks. In the homogeneous networks, the spreading of a dynamical state is a threshold phenomenon: it occupies a valuable fraction of network nodes as the control parameter determining the formation rate of the state exceeds some critical value. Alternatively, in the scalable networks, the dynamical state spreads and persists at whatever value of control parameter. 

Fig.~\ref{fig10}.b presents the distributions of  nodes with periodically changing protein concentrations vs. the periods of  changes  in the undirected scale free graph $\mathbb{G}(10^3,2.2)$ for $\eta=0.1$ (the upper profile) and for $\eta=0.5$ (the lower profile) taken at $a=0.74,$ ${\ }\mu=1$. 
Distributions count all nodes of network which change their protein concentrations periodically disregarding for the amplitudes of these oscillations. Each node has been counted in the distribution just once under the minimal period
of oscillations it exhibits. It is seen that the maximal number of nodes demonstrates the minimal period of changes $\tau=2$ that is not a surprise since the most of negative feedback loops of minimal periods in the scalable networks relays upon the bidirectional edges. 

The statistics of stable long time behavior observed in the scalable regulatory networks can be characterized by the occupancy numbers $p_k$ of the consequent intervals of phase space, $\Delta_k=[T_{k-1},T_k]$. In Fig.~\ref{fig11}, we have shown the occupancy numbers of the model defined on the undirected scale free graph $\mathbb{G}(10^3,2.2)$ with a given configuration of $50$ thresholds u.d. over the unit interval for two opposite values of $\eta.$ Fluctuations shown by the boxes reveal the dependence of the occupancy number upon the certain choice of initial conditions and layouts of switching parameters. It is important to note the high \textit{error tolerance} of scalable regulatory network: in its phase space, the intervals exist which stay void ($p_k=0,$ even though some nodes had  initially been settled into these intervals) and for which $p_k$ is got fixed independently upon the layout of switching parameters and initial conditions.

When the negative interactions present in abundance ($\eta \ll 1$), the valuable fraction of nodes, in the scalable regulatory network, synchronizes  either in the first (next to $x=0$) or in the last (next to $x=1$) intervals of phase space.
The nodes demonstrating oscillations of protein concentrations 
(about $N/2$ total as $\eta\to 0$) are scattered in the middle range of  phase space.

\section{Mean field approach to the large regulatory networks}
\noindent
          
The aim of  present section is to introduce a "mean field"
approach to the large regulatory networks with randomly shuffled 
switching parameters. 

In accordance to (\ref{Th}-\ref{x}), the $j$-th protein synthesis rate   depends upon the exertions $y_i$ of all genes acting on it, that is, the fraction of active arcs incident to $\mathrm{v}_j$ at time $t$. While shuffling randomly the switching parameters in the large regulatory network, we suppose that the positive sign $S=+1$ is selected for the action $i\to j$ with the probability $0<\eta\leq 1$, and the interaction threshold value  equals to some $T_k$ ${\ }(k=1,\ldots, n)$ chosen with some probability $0<P_{ik}\leq 1$, such that $\sum_{k=1}^n P_{ik}=1$ for each gene $i$ acting on $j$. 

The distribution of values $T_k$ over the unit interval can be defined by the set of integrable functions $g_k:[0,1]\to \mathbb{R}^{+}$ satisfying the normalization condition $\int_0^1 g_k(z){\ }dz=1 $ and such that the probability that the threshold value $T_k$ chosen for the interaction does not exceed $x$ reads as 
\[
\mathbb{P}(T_k\leq x)=\int^x_0 g_k(z){\ }dz.
\]
Then the probability that an \textit{outgoing} arc is \textit{active} at time $t$ equals to 
\begin{equation}\label{Lambda}
\Lambda_{\eta}(x^t_i)=\sum_{k=1}^n P_{ik}\left( \eta  \int^{x^t_i}_0 g_k(z){\ }dz +  (1-\eta) \int_{x^t_i}^1 g_k(z){\ }dz\right),
\end{equation}
so that the expected exertion of the gene $j$ at time $t$ is
 \begin{equation}\label{exp}
\langle y^t_j\rangle=\left.\sum_{i=1}^N A^0_{ij}{\ }\Lambda_{\eta}(x^t_i) \right/\sum_{i=1}^N A^0_{ij}.
\end{equation}
The main idea of mean field approach to the large regulatory networks with shuffled switching parameters is to replace the 
exact exertion values $y^t_i$ in (\ref{x}) with their 
expectations (\ref{exp}), 
\begin{equation}\label{x2}
\langle\mathbf{x}^{t+1}\rangle=a\langle \mathbf{x}^t\rangle+(1-a)\langle \mathbf{y}^{t}\rangle,
\end{equation}
in which the expected exertion $\langle \mathbf{y}^{t}\rangle$ is calculated in accordance to (\ref{exp}).

Generally speaking, the problem (\ref{x2}) is as complicated as the original one. However, in some cases it can be more "illustrative" revealing the entire dynamical mechanisms of the system. In the simplest case of uniform probabilities $P_{ik}=1/n$, the system of $N$ coupled equations  (\ref{x2})
is linearized for the uniform distributions, $g_k=1$. After the exponentially fast extinguishing of transient processes, the system (\ref{x2}) with constant $P_{ik}$ and $g_k$ is settled into the uniformly stationary configuration, 
\[
x_{*}{}_i=\frac{(1-\eta)n}{(n+1-2\eta)(N-1)},
\]
independently upon the initial conditions.

For  the "microcanonical" distributions of thresholds, 
$g_k(x)=\delta(x-T_k),$ with the randomly chosen threshold values $T_k$ u.d. over the unit interval and $P_{ik}=1/n$, the dynamics of expectations over the ensemble of large networks ($N\gg 1$) with randomly shuffled switching parameters is given by
\begin{equation}\label{eq}
\langle x_i^{t+1}\rangle=a\langle x_i^t\rangle+(1-a)\left[(1-\eta) f^t(T_m)+\eta\left(1-f^t(T_m)\right) \right]
\end{equation}
where the "mean field" $f^t(T_m)=\left.N(x^t>T_m)\right/N$ is the instantaneous fraction of genes with concentrations $x^t>T_m.$ The "mean" threshold value $T_m$ is determined upon the certain set of randomly chosen thresholds $\{T_k\}$ by the equation (\ref{Lambda}) in the limit $N\gg 1$. In this case, the equation (\ref{eq}) represents a system with a feedback loop. Herewith, depending upon the value of $\eta$ and $T_m$ it can be either positive or negative. The system governed by (\ref{eq})  is getting synchronized very fast, in particular, it exhibits the synchronous oscillations of different periods if $\eta< T_m< 1-\eta$, for $\eta<1/2,$ and $T_m< 1-\eta$, ${\ }T_m>\eta$, 
for $\eta> 1/2.$

In the case of inhomogeneous sets $P_{ik}$, for $N\gg 1$,   the equation (\ref{eq}) can be generalized to
\begin{equation}\label{eq01}
\langle x_i^{t+1}\rangle=a\langle x_i^t\rangle+(1-a)\left[(1-\eta) f^t({T_m}_i)+\eta\left(1-f^t({T_m}_i)\right) \right]
\end{equation}
where ${T_m}_i$ is the local "mean" threshold related to gene $i$. The next step can be done if one supposes some symmetry  properties for the sets of $P_{ik}$ that dramatically reduces the number of independent local "mean" thresholds ${T_m}_i$ to just a few. In such a case, the entire system can be considered as a set of several coupled positive and negative feedback loops whose dynamical behavior can be a subject of detailed analysis. 
In particular, in the system with  several coupled negative feedback circuits characterized by the distinct thresholds ${T_m}_i$, the synchronous oscillations may persist for $\eta>0.$

\section{Conclusion}
\noindent

In the present paper, we have studied the behavior of  large discrete time regulatory networks defined on the homogeneous and scalable maximal graphs. The "traditional" approach to the complex regulatory networks comprising of  a few elements reduces their modelling to the Boolean equations. In this context, the concept of "feedback circuits" has been discussed extensively. However, for the large regulatory networks, a direct logical analysis seems now impossible because of their complexity. 

To get a qualitative insight into the behavior of regulatory networks of thousands elements, we study how sensitive it is upon the shuffling of switching parameters, that are, the interaction signs $S_{ij}=\pm 1$ and the interaction threshold values $T_{ij}$.  Starting from a fixed initial conditions, we have shuffled these switching parameters randomly and then randomized the initial conditions for a given layout of thresholds and interaction signs in the network.

The dynamical behavior observed in such a system crucially depends upon the topology of maximal graph including all possible interactions of genes in the given network. 

We have found that, in the homogeneous regulatory networks, the parameter $0\leq \eta\leq 1$ balancing the positive and negative interactions allowed between genes is a convenient parameter featuring their dynamical behavior. 

When the positive interaction between genes prevail in the system, its dynamical behavior is dominated by the positive feedback circuits responsible for a number of asymptotically stable states, at the same time the influence of negative feedback circuits onto the dynamical behavior of system is negligible. The structure of active subgraph in the regulatory network settled at a fixed point is close to the random graph of Erd\"{o}s and R\'{e}nyi, \cite{ER}. With the increase of percentage of  negative interactions up to approximately $10\%$, the system exhibits a complicated spatiotemporal behavior where the domains of genes with the stationary concentrations of proteins coexist and  interleave with those of  periodically oscillating concentrations.

The feedback circuits formation,  in the regulatory networks defined on the  scale free graphs, relays primarily upon the bidirectional edges connecting  pairs of mutually interacting genes. For the enough fractions of bidirectional edges, in the scalable regulatory networks, oscillations can persist even if there is just a few negative interactions allowed between genes of the network.
 
The model of gene expression regulatory networks which we have considered reveals the similar effect of global scale free topology  on the local dynamics of nodes that has been reported recently in \cite{PS}-\cite{VVB} on the fractions of infected agents in the undirected scalable networks vs. the effective spreading rates $\lambda$. In the homogeneous networks,  a dynamical state  occupies a valuable fraction of network nodes as a control parameter determining the formation rate of the state exceeds some critical value. In contrast, in the scalable networks, the dynamical state spreads and persists at whatever value of  control parameter. 

In the phase space of scalable regulatory networks, the intervals exist which stay void (the occupancy numbers $p_k=0$) and for which $p_k$ is got fixed independently upon the layout of switching parameters and initial conditions. As usual, the scale free networks demonstrate their high error tolerance.

The important unsolved problem deserving a thorough investigation  is that of dynamics of oscillating domains in the graph. Due to the presence of translational invariance (in average) in the homogeneous large regulatory gene expression 
networks, the oscillating domains can probably move in the graph by the interchange of nodes with the stationary part of the network. We can also suggest that perhaps, in the inhomogeneous large regulatory gene expression networks defined on the scale free graphs, the motion of oscillating domains proceeds not in the "real space" of the graph, but rather in between different classes of nodes having the same connectivities. For instance, 
we have beheld that while propagating throughout a scale free graph, the dynamical state occupies at first the nodes with minimal connectivities, then those of maximal connectivities (hubs) and eventually the nodes of intermediate connectivities. Similar dynamical properties of the scale free networks has been reported recently in \cite{VVB}. We shall discuss this issue in the forthcoming publications.

\section{Acknowledgements}
\noindent

One of the authors (D.V.)  benefits from 
la Programme de Bourses de Recherche Scientifique et Technique de l'OTAN CNAM (NATO) (France) that he gratefully acknowledges.
Authors are grateful to L. Streit, F. Russo, M. Grothaus, and all participants of Madeira Math Encounters XXV, Stochastic Analysis (Funchal, Madeira, Portugal) for the interesting discussions on the paper.

\nonumber\section{Appendix A: }
\noindent

The large scale metabolic networks have been studied extensively in \cite{JTAOB}. It has been shown that for many different living organisms the probability degree distributions for both incoming and outgoing edges of a node in the interaction graphs follow the power law, $P(K)\propto K^{-2.2}$, where $K$ is the connectivity of a node. A flexible algorithm  generating scale free graphs based on the principle of evolutionary selection of a common large-scale structure of biological networks incompatible with the preferential attachment approach \cite{BA}, ( see \cite{JMOB}), has been discussed in \cite{VB} recently.
A random procedure which we use to generate the scale free graph
draws back to the "toy" model for a system being at a threshold of stability reported in \cite{FVL}. Here we briefly explain this algorithm for convenience of the readers. 

One consider three random variables $x,y,$ and $z$ that are the
real numbers distributed in accordance to the distributions $f,$
$g$, and $v$ within the unit interval $[0,1].$ We assume that $x$
represents the current performance of a biological network (say,
the protein-protein interaction map), while $y$ and $z$ are the
thresholds for outgoing and incoming edges respectively. The
network  is supposed to be stable until $x<y$ and $x<z$, and is
condemned otherwise. Fluctuations of thresholds reflect the
changes of an environment.

The random process begins on the set of $N$ vertices with no edges
at time $0,$ at a chosen vertex $i$. Given two fixed numbers
$\mu\in[0,1]$ and $\nu\in [0,1]$, the variable $x$ is chosen
with respect to pdf $f$, $y$ is chosen with pdf $g$, and $z$ is
chosen with pdf $v$, we draw $e_{ij}$ edge outgoing from $i$
vertex and entering $j$ vertex if $x<y$ and $x<z$ and continue
the process to time $t=1.$ Otherwise, if $x\geq y$ ($x\geq z$),
the process moves to other vertices having no outgoing (incoming)
links yet.

At time $t\geq 1,$ one of the three events happens:

\textbf{i}) with probability $\mu$, the random variable $x$ is chosen
with pdf $f$ but the thresholds $y$ and $z$ keep their values
they had at time $t-1$.

\textbf{ii}) with probability $1-\mu,$ the random variable $x$ is chosen with
pdf $f,$ and the thresholds $y$ and $z$ are chosen with pdf $g$
and $v$ respectively.

\textbf{iii}) with probability $\nu,$ the random variable $x$ is chosen with pdf
$f$, and the threshold $z$ is chosen with pdf $v$  but the
threshold $y$ keeps the value it had at time $t-1$.

If $x\geq y$, the process stops at $i$ vertex and then starts at
some other vertex having no outgoing edges yet. If $x\geq z,$ the
accepting vertex $j$ is blocked and does not admit any more
incoming link (provided it has any).  If $x<y$ and $x<z$, the
process continues at the same vertex $i$ and goes to time $t+1.$

It has been shown in \cite{VB} that the above model exhibits a
multi-variant behavior depending on the probability distribution
functions $f,$ ${\ }g,$ and $v$ chosen and values of relative
frequencies $\mu$ and $\nu$. In particular, if $\nu=0$, both
thresholds $y$ and $z$ have synchronized dynamics, and sliding
the value of $\mu$ form 0 to 1, one can tune the statistics of
out-degrees and in-degrees simultaneously out from the pure
exponential decay (for $\mu=0$) to the power laws (at $\mu=1$ )
provided $f,$ $g$, and $v$ belong to the class of power law
functions. For instance, by choosing the probability distribution
functions in the following forms
\begin{equation}\label{ss}
  \begin{array}{lc}
   f(u) = (1+\alpha) u^{\alpha}, & \alpha > -1, \\
v(u)\equiv  g(u) = (1+\beta) (1-u)^{\beta},& \beta > -1,
  \end{array}
\end{equation}
one obtains that
\begin{equation}\label{ss01}
P_{\mu=1}(K) \simeq_{K\gg 1} \frac{(1+\beta)\;
\Gamma(2+\beta)\; (1+\alpha)^{-1-\beta}}
{K^{2+\beta}}\; \left( 1+0\left(\frac{1}{K}\right) \right)\;.
\end{equation}
For different values of $\beta$, the exponent of the threshold
distribution, one gets all possible power law decays of
$p_{\mu=1}(K)$. Notice that the exponent $\gamma= 2+\beta$
characterizing the decay of $p_{\mu=1}(K)$ is independent of the
distribution $f(u)$ of the state variable $x$. In the
uncorrelated case, $\mu=0,$ the degree distribution functions
decays exponentially (for instance, $p_{\mu=0}=2^{-K}$ for
$f=g=v=1$) \cite{VB}. For the intermediate values of $\mu$, the
decay rate is mixed.

The preference attachment matrix $\Pi_{SK}$ which elements are the probabilities that a vertex with degree $S$ is connected to a vertex having degree $K$, for a scale free graph, generated in accordance with the above algorithm depend only of one variable $K$, \cite{VVB}, 
\begin{equation}\label{prefVB}
\Pi(K)=(1+\beta)\left(1-\frac K{N-1}\right)^\beta, \quad
\beta>-1.
\end{equation}
Expanding the binomial in the above equation, one gets the
leading term $\propto (K/N-1)^{\beta}$.

\newpage

\section{Figures}
\noindent

\begin{figure}[ht]
 \noindent
 \begin{minipage}[b]{.36\linewidth}
 \begin{flushleft} 
\epsfig{file=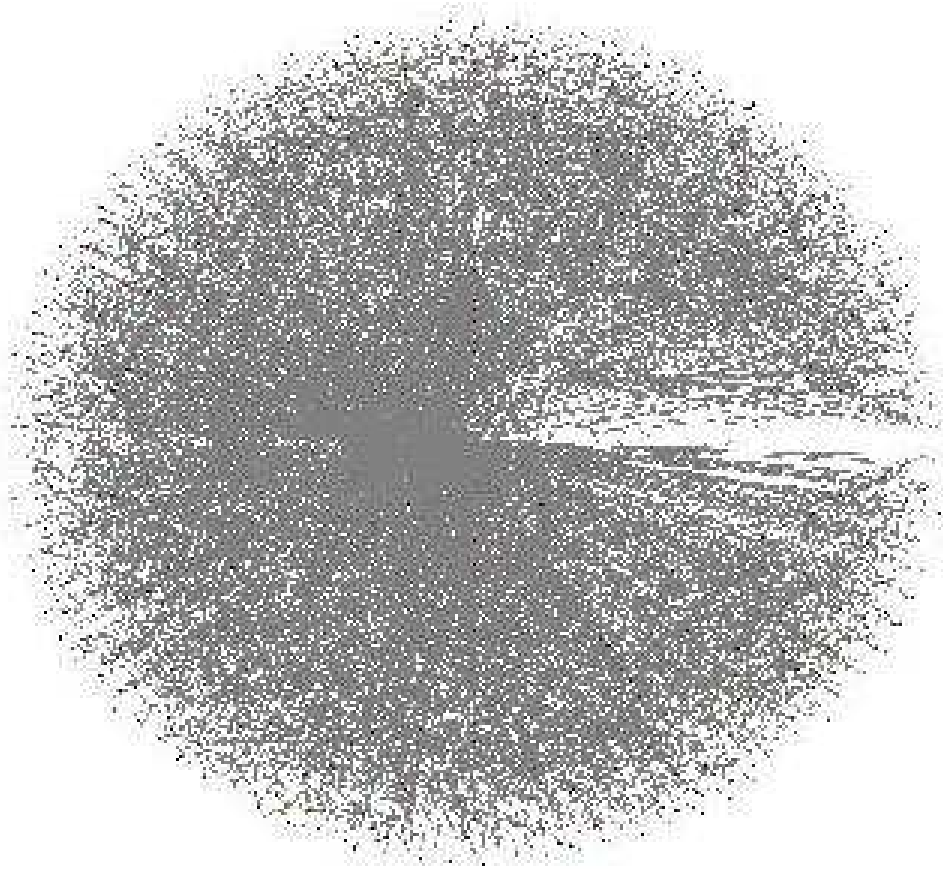, angle= 0,width=13.0cm}
 \end{flushleft}
\end{minipage}
\caption{\label{fig8} The scale free graph  $\mathbb{G}(-2.2,10^3)$ generated in accordance to the algorithm explained in the Appendix A and used for the numerical simulations on the model (\ref{Th}-\ref{x}). It is characterized by the binomial preference attachment $\Pi(K)\sim \left(1-\left.K\right/N-1\right)^{0.2}$. Both probability degree distributions for in-degrees and out-degrees follow the power law $P(K)\propto K^{-2.2}$ that is reported to be typical for the large scale metabolic networks of 43 different organisms studied in \cite{JTAOB}.  On the figure, we have drawn the vertices with higher connectivities  closer to the center of the graph and those of lower connectivities are located on the periphery, so that the nodes equally distanced from the center have equal connectivities. }
\end{figure}   

\begin{figure}[ht]
 \noindent
 \begin{minipage}[b]{.36\linewidth}
 \begin{flushleft} 
\epsfig{file=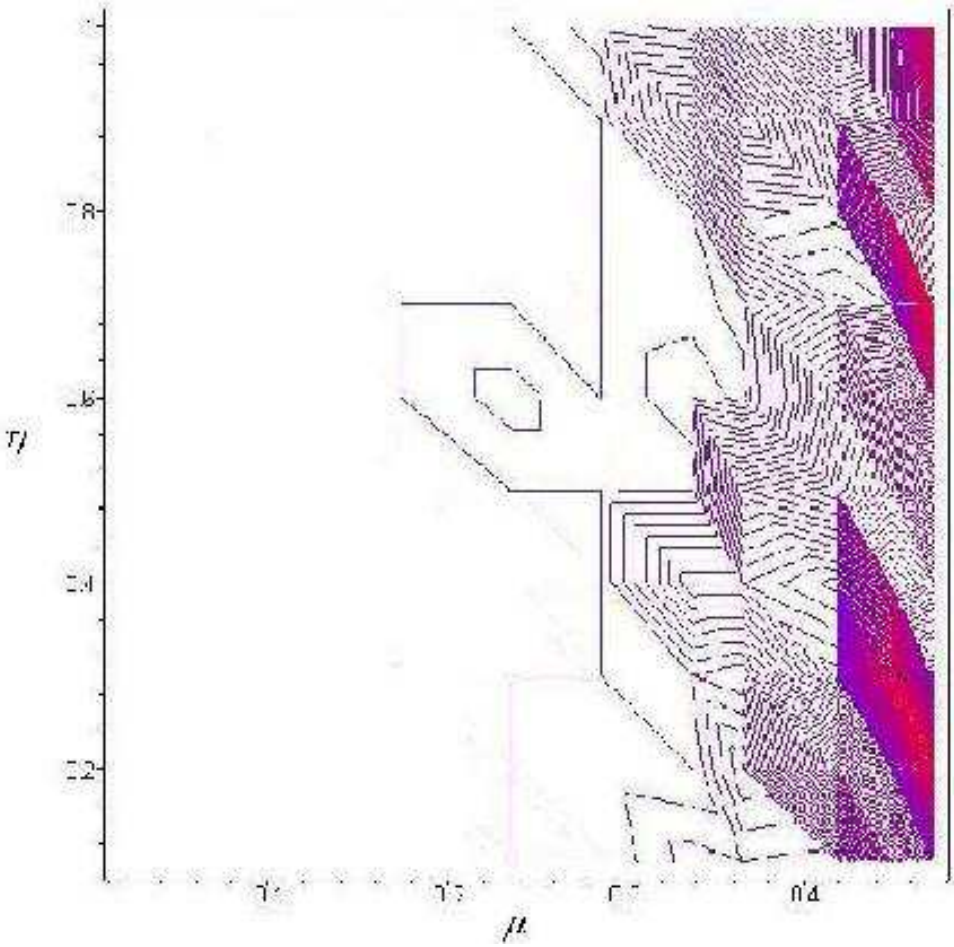, angle= 0,width=13.0cm}
 \end{flushleft}
\end{minipage}
\caption{\label{fig9} The phase diagram $(\eta,\mu)$ displays the appearance of persistent oscillations in the model (\ref{Th}-\ref{x}) defined  on the inhomogeneous scale free graph $\mathbb{G}(10^3,2.2)$ taken at $a=0.3$.  Contours  bound the regions with the equal numbers of oscillating nodes. For small fractions of bidirectional lines $\mu\ll 1,$ oscillations fade out, while they arise when $\mu> 0.2$ persisting at any rate $\eta>0$. The details of the phase diagram strongly depends upon the precise structure of random scale free graph.}
\end{figure}       

\begin{figure}[ht]
 \noindent
 \begin{minipage}[b]{.36\linewidth}
 \begin{flushleft} 
\begin{tabular}{rlrl}
 a. &\epsfig{file=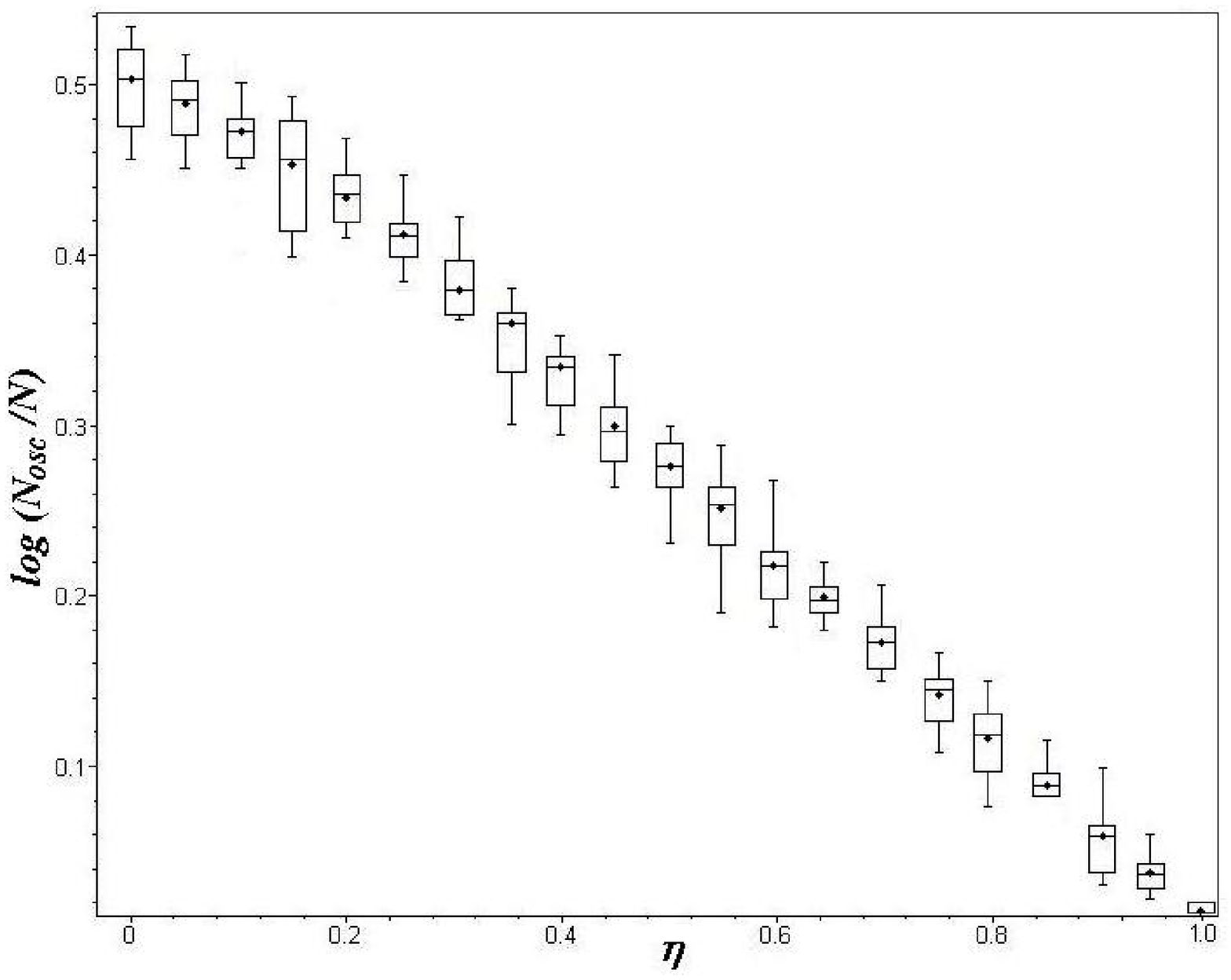, angle= 0,width=7.5cm}&b.&
 \epsfig{file=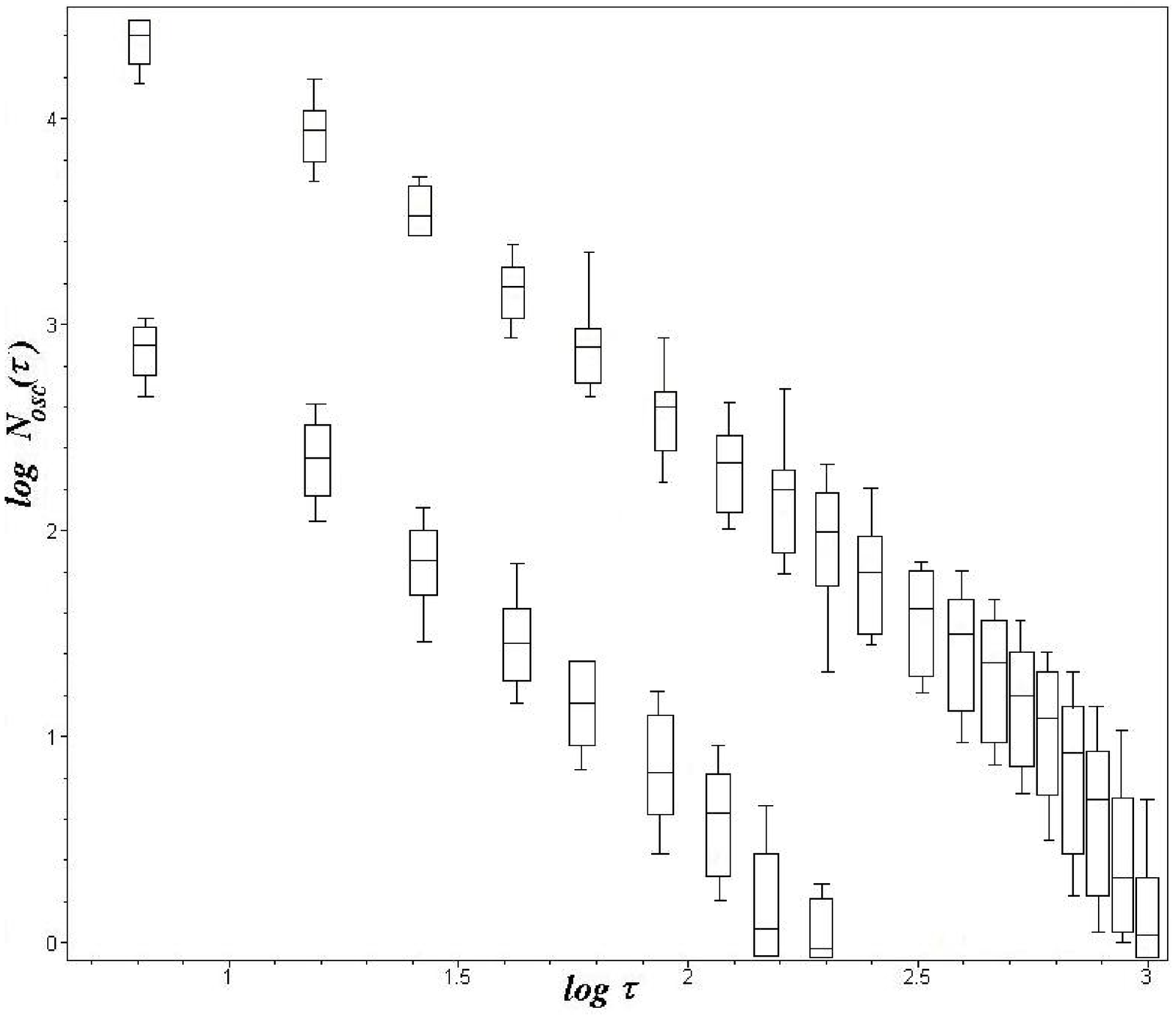, angle= 0,width=7.8cm}
 \end{tabular}
 \end{flushleft}
\end{minipage}
\caption{\label{fig10} a). Decreasing  of oscillating domains  in the regulatory network defined on the undirected scale free graph $\mathbb{G}(10^3,2.2)$ at $\mu =1$.  Boxes present the fluctuations of data collected over the ensemble of $500$ different random layouts of switching parameters and initial conditions, $a=0.74$. The bold points stand for the means. 
b). The distributions of  nodes with oscillating protein concentrations vs. the periods of oscillations  in the undirected scale free graph $\mathbb{G}(10^3,2.2)$ at $\eta=0.1$ (the upper profile) and at $\eta=0.5$ (the lower profile) taken at $a=0.74$ and $\mu =1.$  }
\end{figure}

\begin{figure}[ht]
 \noindent
 \begin{minipage}[b]{.36\linewidth}
 \begin{flushleft} 
\begin{tabular}{rlrl}
 a. &\epsfig{file=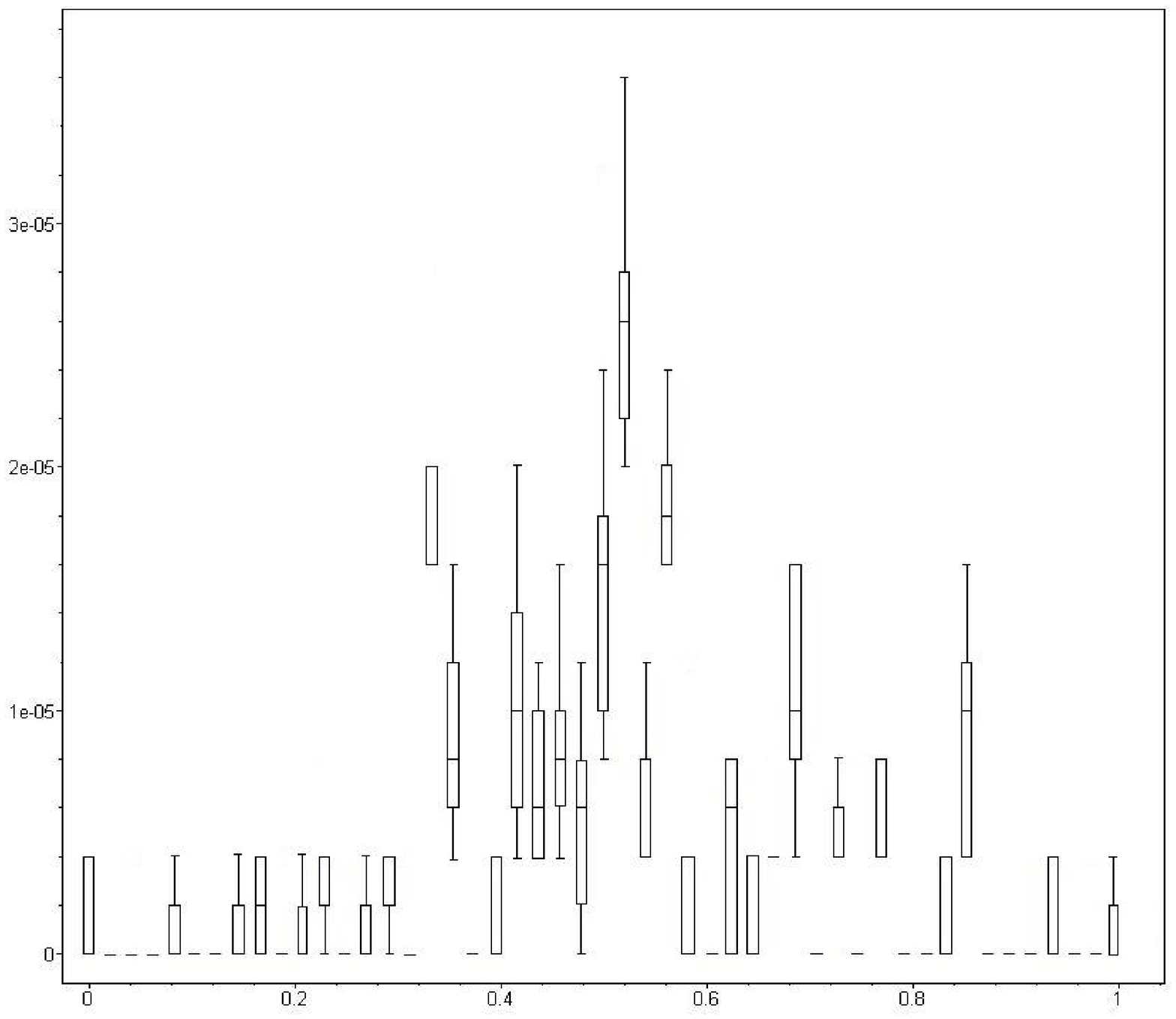, angle= 0,width=7.5cm}&b.&
 \epsfig{file=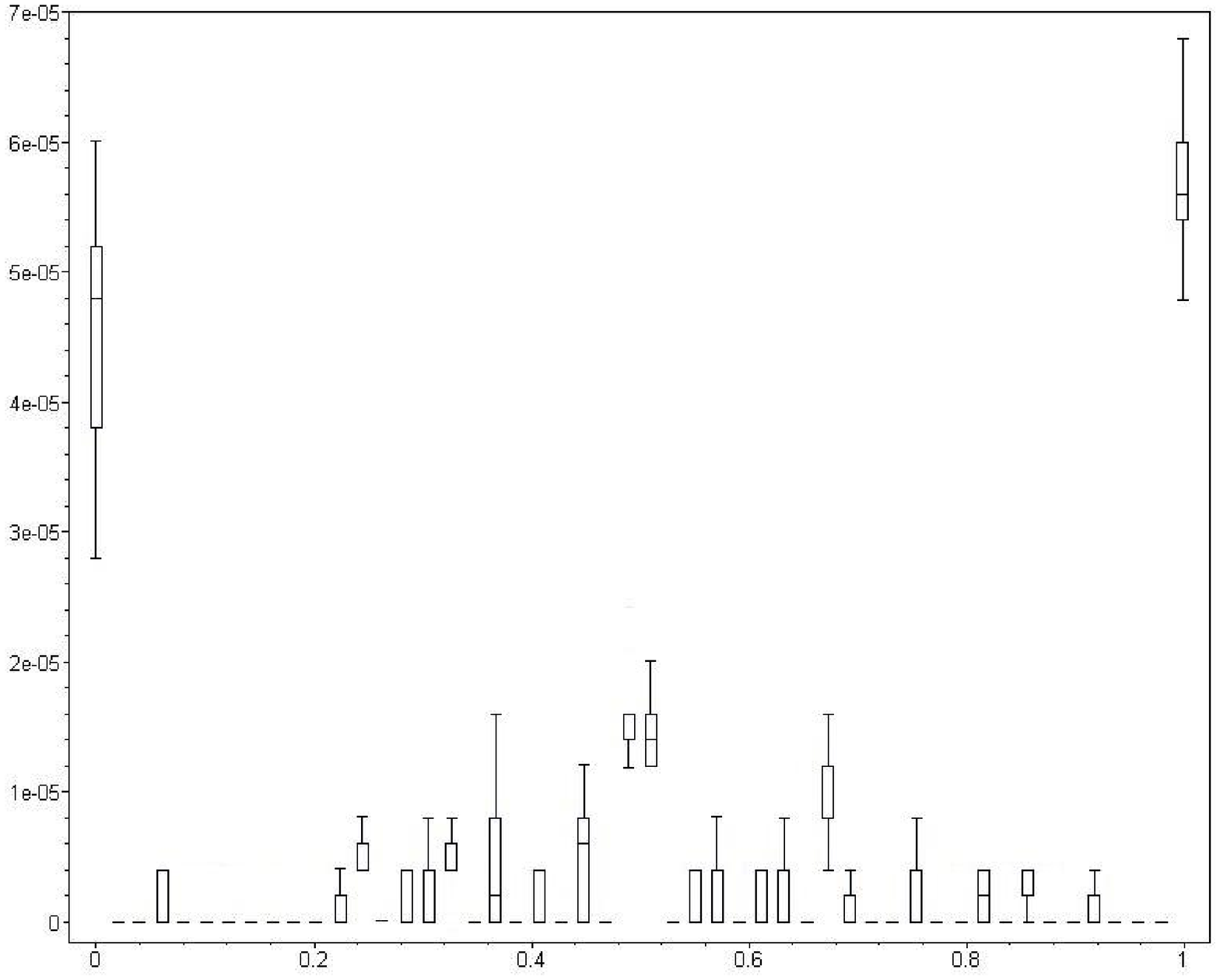, angle= 0,width=7.8cm}
 \end{tabular}
 \end{flushleft}
\end{minipage}
\caption{\label{fig11}  The occupancy numbers $p_k$ of the consequent intervals of phase space, $\Delta_k=[T_{k-1},T_k],$ in the model defined on the undirected scale free graph $\mathbb{G}(10^3,2.2)$ with a given configuration of $50$ thresholds u.d. over the unit interval at $a=0.7$. Fluctuations shown by the boxes reveal the dependence of occupancy numbers upon the certain choice of initial conditions and layouts of switching parameters ($500$ different configurations had been used to collect data). a). for $\eta=0.99$ ($1\%$ of negative interactions),   b). for $\eta=0.01$ ($99\%$ of negative interactions). }
\end{figure}


\begin{thebibliography}{99}


\bibitem{ST}

E. H. Snoussi, R. Thomas, Bulletin of Math. Biol. \textbf{55} (5), 973 (1993).

\bibitem{TK}

R. Thomas, M. Kaufman, Chaos \textbf{11} (1),  180 (2001).

\bibitem{M}

J. D. Murray, \textit{Mathematical Biology} (Springer-Verlag,
Berlin, 1993). 

\bibitem{PS}

R. Pastor-Satorras, A. Vespignani, Phys. Rev. Lett. \textbf{86},
No 14, 3200 (2001).

 \bibitem{VVB}
D. Volchenkov, L. Volchenkova, Ph. Blanchard, Phys. Rev,
{\bf E66} (4), 046137 (2002); Virt. Jour. of Biol. Phys. Res., \textbf{4}(9) (2002), available at texttt{http://www.vjbio.org.}

                    


\bibitem{LFM}
R. Lima, B. Fernandez, A. Meyroneinc, "Modelling the discrete dynamics of genetic regulation networks via real mappings", 
in preparation (2003).

\bibitem{JTAOB}

H. Jeong, B. Tombor, R. Albert, Z. N. Oltvai, A.-L. Barab\'{a}si,
Nature \textbf{407}, 651 (2000).

\bibitem{JMOB}

H. Jeong, S. P. Mason, Z. N. Oltvai, A.-L. Barab\'{a}si, Nature
\textbf{411}, 41 (2001).

\bibitem{T}

D. Thieffry, \textit{Qualitative Analysis of Gene Networks} in the M\'{e}moire pour l'obtention d'Agr\'{e}g\'{e} de l'Eseignement Sup\'{e}rieur, Univ. Libre de Bruxelles (2000).   

\bibitem{Th}

R. Thomas, "On the relation between the logical structure of systems and their ability to generate multiple steady states or sustained oscillations", \textit{Springer Series in Sinergetics}
\textbf{9}, 180 (1981).

\bibitem{PMO}
E. Plahte, T. Mestl, S. Omholt,  \textit{J. Biol.Syst.} \textbf{3}, 1 (1995).

\bibitem{Th94}

R. Thomas, \textit{Ber. Brunzenges. Phys. Chem.} \textbf{98}, 1148 (1994).
\bibitem{TSRT}

D. Thieffry, E.H. Snoussi, J. Richelle, R. Thomas, J. of Biological Systems, \textbf{3} (2), 457 (1995).
 
\bibitem{ER}

P. Erd\"{o}s, A. R\'{e}nyi, Publ. Math. Inst. Hungar. Acad. Sci. \textbf{5}, 17 (1960).
                   
\bibitem{RGBook}
In the random regular graphs $\mathbb{G}(N,r)$, the fixed connectivity $r>3$ insures the presensce of many Hamilton cycles traversing all nodes of the graph, see S. Janson, T. \L uszak, A. Rucinski, {\textsl{Random Graphs}}, John Wiley $\&$ Sons, NY (2000).  
      
\bibitem{Snous}

E. H. Snoussi, \textit{Dyn.Stability Syst.} \textbf{4}, 189 (1989).
             
\bibitem{N2001}

M. E. J. Newman, arXiv:cond-mat/0104209 (2001).

 \bibitem{TR}

D. Thieffry, D. Romero, BioSystems \textbf{50}, 49 (1999).
    

\bibitem{BA}

A.-L. Barab\'{a}si, R. Albert, Science \textbf{286}, 509 (1999).
        

\bibitem{VB}

D. Volchenkov, Ph. Blanchard,  Physica A \textbf{315}, 677 (2002).

\bibitem{FVL}
D. Volchenkov, E. Floriani, R. Lima,  J.Phys.A: Math. and Gen.
\textbf{36}, 4771 (2003). 


\bibitem{CM1}
H. Chat\'{e}, P. Manneville, Chaos {\bf{2}}, 307 (1992).


\bibitem{Chate1}
H. Chat\'{e} and P. Manneville,  Europhys. Lett. {\bf 6}, 591 (1988).

\bibitem{VSBC}
D. Volchenkov,  S. Sequeira, Ph. Blanchard, M.G. Cosenza, Stochastic and Dynamics, \textbf{2}
(2), 203 (2002).
                                


\end{thebibliography}
\end{document}